\documentclass{article}

\usepackage{amssymb,amsfonts,amsmath}

\usepackage{cite,enumerate,float}

\usepackage{color}
\usepackage{tikz}
\usepackage{xcolor}
\usepackage{comment}
\usepackage{cancel}
\usepackage{comment}
\usetikzlibrary{arrows,snakes,backgrounds}
\usepackage{hyperref}

\def\be{\begin{eqnarray}}
\def\ee{\end{eqnarray}}

\def\d{\partial}

\newcommand\bfloor[1]{\lfloor #1 \rfloor}
\newcommand\Lop{\mathcal{L}}
\newcommand\bangle[1]{\langle #1 \rangle}
\renewcommand{\[}{\begin{equation}}
\renewcommand{\]}{\end{equation}}
\renewcommand{\emptyset}{\varnothing}

\definecolor{red}{rgb}{1,0,0}
\definecolor{orange}{rgb}{1,0.5,0}
\definecolor{violet}{rgb}{0.7,0,1}

\begin{document}

\hfill MIPT/TH-09/23

\hfill FIAN/TH-07/23

\hfill ITEP/TH-10/23

\hfill IITP/TH-08/23

\bigskip

\begin{center}
   { \Large{\bf Position space equations for Banana Feynman diagrams

}}

\bigskip

{ V. Mishnyakov$^{a,b,c,e,}$\footnote{mishnyakovvv@gmail.com}, A. Morozov$^{a,d,}$\footnote{morozov@itep.ru},
P.Suprun$^{a,}$\footnote{suprun.pa@phystech.edu}
\\}

\bigskip

    $^a${\em NRC Kurchatov Institute - ITEP, Moscow, Russia} \\
$^b$ {\small {\it MIPT, Dolgoprudny, 141701, Russia}}\\
$^c$ {\small {\it Lebedev Physics Institute, Moscow 119991, Russia}}\\
$^d$ {\small {\it Institute for Information Transmission Problems, Moscow 127994, Russia}}
\\
$^e$ {\small {\it Institute for Theoretical and Mathematical Physics,
Lomonosov Moscow State University, 119991 Moscow, Russia}}

\end{center}

\bigskip
\begin{abstract}
    
The answers for Feynman diagrams satisfy various kinds of differential equations --
which is not a surprise, because they are defined as Gaussian correlators,
possessing a vast variety of Ward identities and superintegrability properties. We study these equations in the simplest example
of banana diagrams.
They contain any number of loops, but can be efficiently handled in position
rather than momentum representation, where loop integrals do not show up.
We derive equations for the case of scalar fields,
explain their origins and drastic simplification at coincident masses.
To further simplify the story we do not consider coincident points,
i.e. ignore delta-function contributions and ultraviolet divergences for the most part.
The equations in this case reduce to homogeneous and have as many solutions as
there are different Green functions --  $2^n$ for $n$ loops in quadratic theory,
what reduces to just $n+1$ for coincident masses, i.e. for a single field. We comment on the recovery of the delta-functions directly from the homogeneous equations and also compare our result with momentum space formulas known in the literature.

\end{abstract}

\section{Introduction}

The study of Feynman integrals has been a hot topic in theoretical quantum field theory for quite a few decades already. For a summary we refer to the following incomplete list of recent reviews
\cite{ Vanhove:2018mto,Rella:2020ivo,Brown:2015fyf,Klausen:2023gui,Kotikov:2021tai,Loebbert:2022nfu} and specifically an exhaustive book \cite{Weinzierl:2022eaz}. The feature of these integrals that keeps appearing again and again in these studies and holds our attention in the present work is the following one.  While the Feynman integrals themselves are rather sophisticated functions (e.g. they are always transcendental), they still can be described as particular solutions to differential equations that are polynomial, hence significantly simpler.

In principle, the origin of this phenomenon could be traced back to the level of string theory\cite{morozov1992string} where it is arguably related to implicit integrability structures \cite{Gerasimov:2000pr,morozov1994integrability}. The algebraic structures on Feynman graphs, discovered by Connes and Kreimer \cite{Kreimer:1997dp,Connes:1999yr,Connes:1999zw,Connes:2000fe} should provide the technical tools to bind these structures to field--theoretical routine. While this is a nice program, it is not what we attempt to do here. Instead we try to bite another corner and derive these differential equations in as straightforward way as possible, for the sake of clarification. To keep the story simple, we focus on the special banana family of graphs --- which already contains diagrams of arbitrary loop number and evaluates to admittedly sophisticated functions.
We find that, at least for this concrete case, the equations in question can be deduced and explained by almost trivial reasoning --- a phenomenon we see as a good sign of tractability of Feynman diagram theory. 

The set of methods more closely related to our considerations are those that come from considering Feynman integrals as periods of some special kinematic spaces. This approach bridges feynmanistics to motivic studies\cite{Marcolli:2009zy,Rej:2009ik,Brown:2015fyf,Bonisch:2021yfw,Morozov:2009kc} and brings an intrinsic hierarchy of complexity on diagrammatic functions: first come dilogarithms, then higher polylogarithmic functions\cite{Bogner:2016qbf,Panzer:2015ida,Duhr:2015rjo,Duhr:2014woa} and their elliptic analogues\cite{Adams:2015pya,Broedel:2018tgw,Remiddi:2017har,Bezuglov:2020tff} etc. The full picture of this hierarchy is not properly established yet; notice however that the analysis of polylogarithmic branch has already turned out remarkably fruitful for amplitudes in $\mathcal{N}=4$ supersymmetric Yang--Mills theory\cite{DelDuca:2009au,Goncharov:2010jf,Drummond:2010mb,Golden:2014xqa}. Another remarkable feature of period approach is appearance of well--known period spaces, for example ones of Calabi--Yau manifolds\cite{Vanhove:2018mto,Bonisch:2021yfw,Bourjaily:2018yfy,Bourjaily:2019hmc}, which a relevant in particular for the banana graphs. 
The active development of the area, combined with the amount of unanswered questions in it makes it reasonable to think these ideas are to become an important part in future streamlined version of quantum field theory.

\begin{figure}[h]
  \begin{center}
    \includegraphics[width=0.5\textwidth]{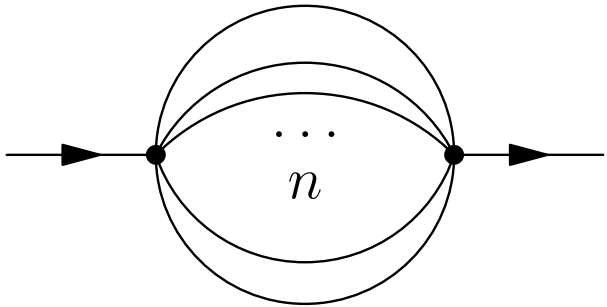}
    \caption{$n$--banana diagram}
    \label{fig:banana}
  \end{center}
\end{figure}

The principal claim of this paper is a proposition that a position--space based approach is no less powerful tool for computation of differential equations on Feynman diagrams than conventional ones, and is significantly simpler conceptually. We demonstrate this for the banana graph family, fig.\ref{fig:banana}. Similar, but less general, considerations can be found in \cite{Lairez:2022zkj,Bonisch:2021yfw}.
The calculations, albeit very simple, and standard in the past \cite{Novikov:1983gd,Morozov:1984goy} are not widely used in current literature. 
The method is applicable in any space-time dimension, since it treats the dimension $D$ as parameter, which enters analytically in the coefficients of the equations. It also works in any number of loops, 
however, the analytic dependence on loop number is not completely deciphered. 
The analyticity with respect to space--time dimension also means all the expressions are dimensionally regularized, which is complemented with the fact that the nonsingular patch in position space does not feel the divergent sectors of momentum space. Together this means the subtleties related to the divergence of momentum space integrals such as scheme (in)dependence, separation of order under integrals etc. can be all avoided. As our basic arguments are fairly general, they probably admit suitable extensions to other classes of diagrams, 
we plan to address this question in our future work.
\\

There are two essential ingredients in the analysis below: 
\begin{itemize}
    \item In position space the banana diagram 
is simply a product of propagators contained in it, 
    \item  These propagators, being Green's functions, are by definition solutions to differential equations:
\[ (\Box + m^2) G = i \delta^{(D)}(x). \]
\end{itemize}
The position space differential equations can then be translated to momentum space by the standard rules of Fourier transforms. Note, that direct Fourier integration of the product of Greens functions requires some substantial work \cite{Groote:2018rpb,Bailey:2008ib}. From this (reciprocal) point of view the key observation is that Fourier transform takes the convolutional banana integral into product in physical space, which is much simpler to handle while doing differential calculations. 

By implementing these ideas, we have computed the differential equations for banana diagrams explicitly, in arbitrary dimension, up to 12 loops. These equations appear not to be random, but rather controlled by some underlying structure; this can be seen empirically by inspecting their coefficients. We present some observations on the structure of these coefficients; sometimes, e.g. in case of leading coefficient, these observations even allow one to predict these coefficients in arbitrary loop number.

The paper is organized as follows. In the first section we specify our computational procedure for the most elaborated case, namely the equal--mass banana diagrams. We describe the computation of the left-- and right--hand sides separately; the right--hand side computation is treated properly for $D=1$ case only. For the left--hand side, we found it instructive to treat $D=1$ case as an example before going into full generality; the reader interested in the procedure only may wish to go to page \pageref{procedure} directly. We also reflect a bit on the nature of the properties of diagram technique we used. In section \ref{sec:Procedure} we list the observations we managed to make on the obtained equations; we also prove some of them. After that we discuss the multiple--mass case in section \ref{sec:DifferentMass}; again, we work the $D=1$ case out before sketching the general scheme. The relation to momentum--space approach and existing results is provided in section \ref{sec:Tomomentum}. We comment on the possible extension of our calculations to other cases in the closing section.  We provide a list of some of the computed equations in the Appendix \ref{Appendix} for illustrative purposes. An even greater list is provided by the supplementing Wolfram Mathematica notebook, accompanying  this paper.

\section{Description of the procedure}\label{sec:Procedure}

In this section we describe the technology under discussion of obtaining differential equations on banana integrals. We found out that the tasks of obtaining homogeneous and inhomogeneous parts of differential equation are better treated separately, for the reason the homogeneous part is controlled by the behavior of diagrammatic functions on big open chart $x \neq y$ where they are smooth, while the inhomogeneous part is a combination of $\delta$--like functions and represents the leap conditions in coordinate origin. Because of that, we first describe the computation of differential operators acting on the function on the left--hand side of our differential equation, and then the computation of inhomogeneous $\delta$--contribution. For the sake of brevity in the next subsection we refer to the homogenized equations $\widehat{\mathcal{L}} G_n = 0$ instead of $\widehat{\mathcal{L}} G_n = \langle\mathrm{deltas}\rangle $ as ``differential equations on banana diagrams'', despite this holds only outside the origin. This, however, will not lead to a confusion, and neither to loss of generality.

\subsection{Computation of differential operator}

Before we go into describing a fairly general method of computing the left--hand side of differential equations under consideration, we need some justification of the assumption that position space differential equations on banana diagrams posses some comprehensible structure. Such a justification is provided by the case of $D=1$ theory, that we work out in the subsection below.

\subsubsection{A toy case: $D=1$}

The question of defining banana integrals by differential equations is the most simple in one spacetime dimension. The reason for it is a particularly simple form of Green's function: in one spacetime dimension it is given by
\[ G(x,y) = \frac{\sin |m (x - y)|}{m} \]
While all the equations obtained below can be deduced directly from equations of motion $(\d^2 + m^2) G = 0$ outside of $x = 0$, here we rely on the explicit form of Green function instead; this allows both simpler and more conceptual derivation and a general--$n$ answer.

In principle, it is possible to derive the operator of interest by direct comparison of $\sin^n (m x)$ functions with their derivatives, but we found it more enlightening to express non--singular patch of Green's function in terms of exponentials, that is, by its Fourier harmonics:
\[ G(x) = \frac{1}{2 i m} e^{i m x} + \frac{-1}{2 i m} e^{- i m x} \]
$n$--banana diagram, being a position space product of propagators, is therefore a linear combination of several exponents, coming out as products of harmonics of individual propagators:
\[ G_n(x) =  G^n (x) = \left(\frac{1}{2 i m} e^{i m x} + \frac{-1}{2 i m} e^{- i m x}\right)^n = \alpha_{-n} e^{-i n m x} + \alpha_{-n+2} e^{ - i (n-2) m x} + \ldots + \alpha_n e^{i n m x} \]
On the other hand, each exponential contribution obeys a defining differential equation $ (\d - i a) e^{i a x} = 0$. Therefore, the product of such operators will kill each exponent in the expansion:
\[\label{D1eqm} (\d + i n m) (\d + i (n-2) m) (\d + i (n-4) m) \ldots (\d - i n m) G_n (x) = \prod\limits_{\substack{k=-n\\ k\equiv n\, \operatorname{mod} \, 2}}^{n} (\d + i k m) G_n = 0 \]
\textbf{This is the differential equation we were to obtain}. However, for the purpose of comparison with higher--dimensional results we express it in terms of universally meaningful operators $\Box = \d^2/\d x^2$ (i.e., Laplace operator) and $\Lambda = x \frac{\d}{\d x}$ (i.e., dilatation operator $x^\mu \d_\mu$):

for odd $n$:
\[\label{D1odd} \prod\limits_{k=0}^{\bfloor{\frac{n}{2}}} (\Box + (n-2 k)^2 m^2) G_n = 0 \]

for even $n$:
\[\label{D1even} \Lambda \prod\limits_{k=0}^{\frac{n}{2} - 1} (\Box + (n - 2 k)^2 m^2) G_n = 0 \]
Notice that these equations are written for arbitrary number of loops, and have a simple factorized form.

\subsubsection{General $D$}\label{sectionGeneralD}

Now we are going to obtain equations similar to provided above for arbitrary $D$. In general dimension, no simple formula for Green's function exist (notice the special case of $D=3$, see below), and no $D=1$--like argument with Fourier expansion passes. For the reason, we have to compute the equations from differential--algebraic manipulations rather then some byside trick. However, here a problem is faced: a single equation from the definition of Green's function
\[ (\Box + m^2) G_1 = 0 \]
is no more enough to fix a function of several variables. However, we can get additional conditions from the Lorenz--invariance of Green's function:
\[ x_{[\mu} \d_{\nu ]} G = (x_\mu \d_\nu - x_\nu \d_\mu ) G = 0\]
Consequently, a question arises: do Lorenz--invariance conditions combined with shell condition provide enough differential relations to close down the relations between derivatives of arbitrary power of propagator? The answer is actually yes, as can be seen on the following

\paragraph{Example.}

  Take $n=2$ banana diagram: $G_2 = G^2$. By successive differentiating it, we obtain

  \[\label{n2diff} \begin{aligned}
    \d_\mu G_2 &= 2 G \d_\mu G \\
    \d^\mu \d_\mu G_2 &= 2 G \Box G + 2 (\d_\mu G)^2 = - 2 m^2 G^2 + 2 (\d_\mu G)^2
  \end{aligned} \]
  For the reason, $ (\Box + 2m^2) G_2 = 2(\d_\mu G)^2 $ and what remains is to kill $ (\d_\mu G)^2 $. To do it, an obvious option is to apply one extra differentiation:

  \[\label{n2diff2} \begin{split} x^\nu \d_\nu (\d_\mu G)^2 &= 2 (x^\nu \d_\nu \d_\mu G) (\d_\mu G) = \\
     &= - 2 D (\d_\mu G)^2 + 2 (\d_\nu x^\nu \d_\mu G) (\d_\mu G) = - 2 D (\d_\mu G)^2 + 2 (\d_\nu x^\mu \d_\nu G) (\d_\mu G) =\\
     &= - 2 D (\d_\mu G)^2 + 2 (\d_\mu G)^2 - 2 x^\mu m^2 G (\d_\mu G) = - (2D-2) (\d_\mu G)^2 - m^2 (x^\mu \d_\mu) (G^2)  \end{split} \]
  where successively used rules of differentiation, commutation relation of $x^\nu$ and $\d_\nu$, Lorenz invariance and Leibniz rule again. Gathering all the steps together, we deduce that 2--banana diagram obeys the following differential equation:

  \[\label{G2eqD} (x\d) (\Box + 4 m^2 ) G_2 + (2D - 2)(\Box + 2 m^2) G_2 = 0 \ \blacksquare \]
  In this example, we used the Lorenz invariance conditions directly, in form of differential relations. While this makes the derivation above more clean, for larger $n$ we found it more practical to exploit them in form of $G = G(x)$ where $x = \sqrt{x^\mu x_\mu}$, i.e. change the argument in banana functions and treat them as functions of one variable (the magnitude of point separation), keeping the treatment essentially one--dimensional. The important change from the $D=1$ case is that the shell condition is now not $(\d^2 + m^2)G = 0 $ but rather

  \[ \left(\frac{\d^2}{\d x^2} + \frac{D-1}{x} \frac{\d}{\d x} + m^2\right) G = 0 \]
  (cf. Laplace operator in spherical coordinates). We'll see that this affects our results significantly. For the rest of this section $\d$ without indices will stand for $\frac{\d}{\d x}$, for $x$ defined above.

  From the above example it can be seen that the principal difficulty in constraining the powers of propagator $G(x)^n$ with a differential equation is the emergence of products of derivatives $G^k (\d G)^l$ (there presented as a function $(\d_\mu G)^2$) after differentiating the banana function. To keep track of them we introduce a special notation

  \[ I_k^{(n)} = (n)_k G^{n-k} (\Lambda G)^k \]
  where $(n)_k = n (n-1) (n-2) \ldots (n-k+1) $ is the falling Pochhammer's symbol with $k$ brackets, which were introduced for later convenience, and we chose to pack all the first--order derivatives into $\Lambda$--operators. Notice that the upper index depends on the the order of banana only. The derivatives of $I_k^{(n)}$'s can be expressed back through $I_k^{(n)}$'s by the following differential identities:

  \[
  \begin{split}
    \Lambda I^{(n)}_k &= (n)_k (n-k) G^{n-k-1} (\Lambda G)^{k+1} + (n)_k k G^{n-k} (\Lambda G)^{k-1} (x^2 \d^2 G + x \d G) =\\
    &=(n)_{k+1} G^{n-k-1} (\Lambda G)^{k+1} + (n)_k k G^{n-k} (\Lambda G)^{k-1} (- (D-2)  (\Lambda G) - x^2 m^2 G ) = \\
    &=I^{(n)}_{k+1} - k (D-2) I^{(n)}_k - k x^2 (n-k+1) m^2 I^{(n)}_{k-1}
  \end{split}
  \]
  where on the second line we used the equation on single Green's function, and then restored the definition of $I_k^{(n)}$. The last line allows one to express $I_{k+1}^{(n)}$ in terms of $I_{l}^{(n)}$'s with lower $l$ and their derivatives. Recursively applying such identities, one can express $I_k^{n}$ for arbitrary $k < n$ in derivatives of $I_1^{n}$ and $I_0^{n}$. As we have

  \[
  \begin{split}
    I_0^{n} &= G^n = G_n\\
    I^{(n)}_{1} &= n G^{n-1}\Lambda G= \Lambda G^{n}
  \end{split}
  \]

this is expression of $I_k^{(n)}$ in terms of $n$--banana function and its derivatives. However, the derivative of $I_{n}^{n}$ does not involve any further contributions:

  \[ \Lambda I_n^{(n)} = \Lambda (\Lambda G)^n = n (\Lambda G)^{n-1} (\Lambda^2 G) = - n (D-2) I^{(n)}_n - n x^2  m^2 I^{(n)}_{k-1} \]
  (this is the above relation with $k$ set equal to $n$: the falling factorial in $I^{(n)}_{k+1}$ vanishes). Therefore, if we substitute in this relation $I_n^{(n)}$ and $I_{n-1}^{(n)}$ with their expressions in terms of $n$--banana function obtained before, then we arrive on a differential equation on $G_n$.

  It is easy to see that our procedure sequentially expresses the $I_k^{(n)}$--functions as linear combinations of derivatives of banana function, i.e., as the results of action of linear differential operators on it. For the sake of clarity we would like to reformulate the procedure in terms of these operators, as accepted in differential Galois theory under the name of symmetric power of an operator. Denote

  \[ L_k^{(n)} = \text{linear differential operator that maps $G_n$ to $I_k^{(n)}$}\]
  Then the recursion relations we rely on can be restated as

  \[ \label{rec1} \Lambda L^{(n)}_{k} = L_{k+1}^n - k (D-2) L^{(n)}_k - x^2 m^2 k (n-k+1)  L^{(n)}_{k-1} \]
  with initial conditions

  \[
  \begin{split}
    L_0^{(n)} = \mathrm{id}\\
    L_1^{(n)} = \Lambda
  \end{split}
  \]

  \label{procedure} In short, \textbf{our procedure of computation the differential operator in an equation on $n$--banana can be formulated the following way}. Suppose we look for a differential equation for $n$--banana. Then

  \begin{enumerate}

  \item set $L_0 = 1$, $L_1 = \Lambda$

  \item for $k = 2, 3, \ldots n+1$ set $L_k = \Lambda L_{k-1} + (k-1)(D-2) L_{k-1} + x^2 (k-1) (n - k +2)m^2 L_{k-2}$ expanding them recursively with accordance to the rules of differential operator algebra

  \item then $L_{n+1}$ will be the operator on left--hand side of the differential equation on $f$ satisfied by banana diagrams, i.e. kill $G_n$--functions.

  \end{enumerate}

  The outcome of this procedure is an equation on $G_n$ expressed in terms of $x$ and $\Lambda$. However, we found it useful to introduce the $\Box$--operators as well, according to the formula $\Lambda^2 = x^2 \Box - (D-2) \Lambda$. The existence of two differential operators in the same variable introduced additional ambiguities in the form of equation, for the aforementioned relation; moreover, the ordering of operators is also a matter of convention, as they do not commute.

  We voluntaristically resolve these ambiguities by the following conventions: all the $\Lambda$--operators except possibly one are to be replaced $\Box$ with necessary introduction of the lower--order corrections, and we order each term of our operators according to the pattern

  \[ \label{order} x^a \Lambda^{0 | 1} (c_b \Box^b + c_{b-1} m^2 \Box^{b-1} + \ldots) \]
  i.e. first $x$, then no more than one $\Lambda$, and then the $\Box$--polynomial, reading left--to--right. Notice that these ordering conventions are term--by--term, and the terms are selected according to the power of $x$.

  Here we provide some examples of differential equations obtained by application of this procedure. A more complete list may be found in Appendix.

\subsubsection{Remark: $D=3$}
The simplification of the equations obtained by fairly general procedure in concrete dimensions is not bound to $D=1$ case only. For example, it is possible to derive general--$n$ answer for $D=3$ as well. The essential points for it are relations between Green's functions in different dimensions and the general--$n$ answer for $D=1$. In concrete, the three--dimensional Green function is
\begin{equation}
	\left. G \right|_{D=3}= \dfrac{\exp{ \left( \pm i m x \right) }}{x}
\end{equation}
Therefore $x \left. G \right|_{D=3}$ is one--dimensional Green function, so $x^n \left. G_n \right|_{D=3}$ is one--dimensional $n$--banana and as such obeys eq. \eqref{D1odd},\eqref{D1even}. Therefore the general--$n$ answer in $D=3$ is
\[ \prod\limits_{\substack{k=-n\\ k\equiv n\, mod\, 2}}^{n} (\d + i k m) (x^n G_n) = 0.\]
However, unlike $D=1$ case, in three dimensions $\d^2$ is no more Laplace operator, but
\begin{equation}\label{Boxlambda}
	\partial^2= \Box - \dfrac{D-1}{x^2}\Lambda \left. = \right|_{D=3} \Box - \dfrac{2}{x^2}\Lambda
\end{equation}
instead. The equation the form \eqref{order} is obtained by repeated application of this identity and commutation relations.

For example for $n=2$ the equation reads:
\begin{equation}\label{D31}
	\Lambda \left( \partial^2+4m^2 \right) x^2  G_2\big|_{D=3} = 0.
\end{equation}
By \eqref{Boxlambda} we then obtain:
\begin{equation}
	\left( x^2 \Lambda (\Box+4m^2) + 4x^2 \left(\Box+2m^2 \right)  \right)G_2=0
\end{equation}
which is indeed \eqref{G2eqD} at $D=3$.

This calculation can be generalized to any $n$ and will affect the structures appearing in the equations for arbitrary $D$, see \ref{leadproof}. A key advantage of this approach is that it reduces the equations to the basic $D=1$ case and then provides a way to implement corrections related to higher dimensions. For arbitrary odd $D$ the analogous relation is:
\begin{equation}
	G(x) = \int e^{ip x\cos\theta} \delta(p^2-m^2)p^{D-1}dp (\sin\theta)^{D-2} d\theta
	\sim  (\partial^2+m^2)^{\frac{D-3}{2}}\frac{\sin m x}{mx}
\end{equation}
This representation, however, does not immediately provide a simple way to obtain higher differential operators, unlike for $D=3$. We leave the development of this approach for future studies.

\subsection{Computation of the right--hand side}

In this section we present some preliminary considerations on how to compute the right--hand side of differential equations on banana functions. The key ingredient is an observation that the inhomogeneous part is $\delta$--like singular and, in fact, concentrated at one point. We solve the problem completely in $D=1$.

As the right--hand side is purely $\delta$--like we can treat it merely as a bookkeeping device for Green's function discontinuities at the coinciding points, and, consequently, compute it from the local considerations. From the defining identity

\[ \frac{\d}{\d x} \alpha \langle \text{unit jump}_0 \rangle = \alpha \delta(x)\]

we deduce

\[ \label{jumpdelta} \frac{\d^n}{\d x^n} f(x) = (\Delta f) \delta^{(n-1)}(x) + (\Delta f') \delta^{(n-2)}(x) + \ldots + (\Delta f^{(n-1)})\delta (x) + \phi_{reg}(x)\]

(we denote $\Delta u = u(+0) - u(-0)$ and assume that $f$ is smooth for $x \neq 0$ and has no $\delta$--like singularities at $x=0$, but only jump discontinuities; $\phi_{reg}(x)$ is not assumed to be continuous at $0$, but also free from concentrated singularities only). So we see that the singular part of $n$'th derivative of a discontinuous function depends on the jumps of the first $n-1$ derivatives only.

For banana functions, the derivative jumps can be computed from the jumps of the propagator by repeated application of the simple identities
\[\begin{split} \label{jumprec}
&\Delta(A+B) = \Delta A + \Delta B\\
&\Delta(AB) = (\Delta A)\bangle{B} + \bangle{A}(\Delta B)\\
&\bangle{A+B} = \bangle{A} + \bangle{B}\\
&\bangle{AB} = \bangle{A}\bangle{B} + \frac{1}{4}(\Delta A)(\Delta B)\\
\end{split}\]
where $\bangle{u} = \frac{u(+0) + u(-0)}{2}$, as well as the off--$0$ Leibniz rule and shell condition
\[\begin{split} \label{regdiff}
&\d (A B)(\pm 0) = \left. ((\d A) B + A (\d B))\right|_{\pm 0}\\
&\d^n G (\pm 0)= - m^2 \d^{n-2} G (\pm 0)\ (\text{in }D=1)\\
\end{split}\]
which do hold under $\bangle{\ldots}$ and $\Delta(\ldots)$, as these operators are linear in limits $\lim\limits_{x \rightarrow \pm 0}$ which are taken over regularity domains $x > 0$ and $x < 0$.

The initial conditions for this recursive computation depend on a choice of homogeneous part of the propagator. For example, for causal (Feynman) Green's function $\frac{1}{2m}\exp^{i m |x|}$ we have
\[\label{init} \Delta G = 0,\ \Delta(\d G) = 1,\ \bangle{G} = \frac{1}{2m},\ \bangle{\d G} = 0\]

Together, the formulas \eqref{jumprec}, \eqref{regdiff}, \eqref{init} allow one to compute any derivative jump $\Delta(\d^k G_n)$ of a banana function. Now, with derivative jumps at hand, we apply the formula \eqref{jumpdelta} to the linear differential equation \eqref{D1eqm} term--by--term:
\[ \sum c_n \d^n G_n = \sum c_n \sum\limits_{k=0}^{n} (\Delta (\d^k G_n))\delta^{(n-k-1)} \]
and sum all the contributions from individual terms. This completes our procedure for computation of the right--hand side.
\\
\\
Now, the above--described technique produces a differential equation on $n$--banana diagram in $D=1$ of the form
\[ \prod\limits_{\substack{k=-n\\ k\equiv n\, mod\, 2}}^{n} (\d + i k m) G_n = \sum_{k=0}^{n-1} a_k \delta^{(k)}(x)\]
Notice that as the worst discontinuity of $G(x)$ is a kink and not a jump ($\Delta G = 0$, $\Delta(\d G) \neq 0$), the sum on the right--hand side contains the derivatives of $\delta$--function on higher than $\delta^{(n-1)}(x)$. However, for compatibility with higher--$n$ results it is necessary to consider modified equation (see \ref{xdeg})
\[ x^{n-1}\prod\limits_{\substack{k=-n\\ k\equiv n\, mod\, 2}}^{n} (\d + i k m) G_n = x^{n-1} \sum_{k=0}^{n-1} a_k \delta^{(k)}(x)\]
But for the distributional identity
\[ x^{n} \delta^{(k)}(x) = \begin{cases} (-1)^n \frac{k!}{(k-n)!} \delta^{(k-n)} (x)&\ n \leqslant k\\ 0 &\ n > k\end{cases}\]
only the worst--singularity term $ a_{n-1} \delta^{(n-1)}(x)$ survives and the equation takes the form
\[ x^{n-1}\prod\limits_{\substack{k=-n\\ k\equiv n\, mod\, 2}}^{n} (\d + i k m) G_n = A_n \delta(x) \]
The value of $A_n$ can be obtained by the direct computation
\[ A_n = i \frac{n!}{(2 m)^{n-1}}\]
Summing up, \textbf{the complete differential equation on $n$--banana diagram is given by}
\[\begin{split} \label{D1inhom}
& x^{n-1} \prod\limits_{k=0}^{\bfloor{\frac{n}{2}}} (\Box + (n-2 k)^2 m^2) G_n = i \frac{n!}{(2 m)^{n-1}} \delta(x),\ n\text{ odd;}\\
& x^{n-2} \Lambda \prod\limits_{k=0}^{\frac{n}{2} - 1} (\Box + (n - 2 k)^2 m^2) G_n = i \frac{n!}{(2 m)^{n-1}} \delta(x),\ n\text{ even.}\\
\end{split}\]

We note that if one multiplies the equations on an extra power of $x$, the $\delta$--part gets completely eliminated and the equation becomes homogeneous; however, it still possesses solutions that are singular at $0$, due to vanishing coefficient before highest derivative. Such ``homogenized'' equations are less informative than \eqref{D1inhom}, as the derivative discontinuity is not fixed by right--hand side. This also implies that extra carefulness is required while manipulating differential equations on potentially discontinuous functions, such as ours: one can no more cancel extra $x$ power, as $\frac{0}{x} = \alpha \delta(x) \neq 0$ ($\alpha$ arbitrary)!

It is important to find a suitable generalization of these considerations for arbitrary $D$. A tempting opportunity is to take a pullback from the radial coordinate
\[ \delta^{(D)}(x^\mu) = \frac{1}{\mathrm{vol}(S^n) |x|^{D-1}} \delta(|x|)\]
in the full--space (as opposed to radially reduced) shell condition $(\Box + m^2) G = i \delta^{(D)}(x^\mu)$ and then proceed as above. But as the relations between radial and $D$--dimensional $\delta$--functions are delicate, we leave their clarification for future research.

\subsection{Discussion}

\subsubsection{Remark on the origin of equations}\label{demagogy}

The success of above--described operation depends heavily on whether or not differential equations on multipliers and Leibniz rule generate enough constraints to bind the space of products and their derivatives to finite dimension. For further considerations, it is important to know if they do so in general case, or this is the specifics of Green's functions. We argue that this phenomenon is actually general, for the reasoning below.

If $f$ and $g$ are solutions to order--$n$ and order--$m$ linear differential equations, respectively; this means $f^{(n)}$ (resp. $g^{(m)}$ can be expressed as a linear combination of lower derivatives, and by further differentiation and reapplying this identity the same is true for every higher derivative. Therefore any $n+1$ derivatives of $f$ (resp. $m+1$ derivatives of $g$) are linearly dependent. By multiplying the linear relations, we deduce the same holds for any $n+1$ functions $(\d^i f) (\d^k g)$, for fixed $k$ (resp. $m+1$ functions $ (\d^l f) (\d^j g) $, for fixed $l$). Consequently, the dimension of space spanned by functions $ (\d^i f) (\d^j g) $ is no greater than $mn$; but for the Leibniz rule, this space is closed under taking derivatives, so $f g$ has no more than $m n$ linearly independent derivatives and as such obeys an order $mn$--differential equation. This number cannot be lowered provided than we require the equation to kill the product of \emph{any two} solutions of $f$ and $g$ equations ($f_i g_j$).

This line of thought admits a nice geometric interpretation. A function obeying a differential equation together with its derivatives can be thought to span a vector bundle over its domain; the differential equation defines than a flat connection on this bundle, for which $f$ is a flat section. Connections on vector bundles are known to define a connection on their tensor product, given by the formula $ \nabla_{\otimes} (f \otimes g) = (\nabla_f f) \otimes g + f \otimes (\nabla_g g) $, that is flat if the multipliers are flat. As the rank of the tensor product of rank--$n$ and rank--$m$ bundles is $mn$, than by reversing the logic we see that $f \otimes g$ obeys an order--$mn$ differential equation. In our case Lorenz invariance conditions imply that all the functions on consideration are functions of $|x|$ only, therefore the base is one--dimensional, so all connections are flat and this logic does apply.

\subsubsection{Remark on maximal cuts}\label{sec:cuts}
In this section we have manipulated with solutions to homogeneous and inhomogeneos equations. Let us make a more detailed comment on the difference between the two their relation to notion of a maximal cut.
\\\\
To do this we make use of momentum space. A generic solution to the homogeneous Klein-Gordon equation is given by the two momentum integrals:
\begin{equation}\label{G+-}
	\begin{split}
	G_{+}(x)&= \int d^{D}p \  \theta(p^0) \delta(p^2-m^2) e^{i p_\mu x^\mu}
	\\
		G_{-}(x)&= \int d^{D}p \ \theta(-p^0) \delta(p^2-m^2) e^{i p_\mu x^\mu}
	\end{split}
\end{equation}
The difference is in the choice of branch of the square root in the delta function. The solutions of the inhomogeneous equations are given by:
\begin{equation}
	G(x) \simeq \int \dfrac{d^D p }{p^2-m^2}
\end{equation}
where the choice of integration contour should be specified to obtain a specific solution. The solution is given up to a linear combination of $G_{+}(x)$ and $G_{-}(x)$ which correspond exactly to a change in the integration contour. When dealing with Feynman integrals in flat space in a stationary situation one considers the so called causal or Feynman propagator which is a specific solution given by:
\begin{equation}\label{Gc}
	G_c(x)=\int \dfrac{d^Dp}{p^2-m^2+i \epsilon} = \theta(x^0) G_{+}(x)+ \theta(-x^0)G_{-}(x)
\end{equation}
The relation between homogeneous and inhomogeneous equations is usually formulated using the notion of the maximal cut. The prescription is straightforward: in each propagator one substitutes $\frac{1}{p^2-m^2}$ with $\delta(p^2-m^2)$, which is nothing but passing to the imaginary part of each propagator, as can be easily seen from \eqref{Gc}. Clearly, this relation is the same one, that we explore here. The maximal cut is just one of the many solutions to the equation for the banana graphs, which correspond to choosing only a special combination of $G_{+}$ and $G_{-}$ for each line:
\begin{equation}
\begin{split}
	G_{\text{cut}}=G_{+}+G_{-}. ?
	\\
	\left(G_c \right)^n \longrightarrow \left(G_{\text{cut}} \right)^n
\end{split}
\end{equation}
It is distinguished by it's analytical properties, which we do not discuss here. The equations for the banana graphs and it's maximal cut share the same differential operator and differ by the inhomogeneous delta term as calculated above:
\begin{equation}
	\begin{split}
			\mathcal{L}(n) \left(G_{\text{cut}} \right)^n &=0 \\
			\mathcal{L}(n) \left(G_{c} \right)^n &= i \dfrac{n!}{\left(2m\right)^{n-1}}\delta(x)
	\end{split}
\end{equation}
Another comment is due about the dependence of the r.h.s on the choice of representative of the solution to the inhomogeneous equations. Mainly, choosing different combinations of $G_{+}$ and $G_{-}$ changes the discontinuities $\Delta G$ and for the derivatives. The r.h.s in the above subsection where computed for the specific choice of the Feynman propagator. Changing the representative to the, say, retarded Greens function $G_{\text{ret}}=\theta(x^0)G(x)$ would also change the r.h.s.
\\\\
All of the discussed peculiarities do not affect the resulting equations in a crucial manner, therefore we do not further elaborate on that issue. By this discussion we have simply elaborated that the obtained homogeneous equation are nothing but the maximal cut equations discussed in the literature \cite{Lairez:2022zkj,Bonisch:2021yfw}.
\section{Observations on differential equations}\label{sec:Observations}

In this section, we outline the properties of our differential equations we managed to observe. We also sketch the proofs of those of them that we managed to make. In our description of the properties found, we follow the above--described conventions, namely, we write each differential operator in a form
\[ \mathcal{L}(n) = \sum x^a \Lambda^{0 | 1 } (c_b(D) \Box^b + c_{b-1}(D) m^2 \Box^{b-1} + \ldots) \]
where by $\Lop(n)$ a differential operator from the equation on $n$--banana is denoted. We refer to each term in this sum (selected according to the powers of $x$ and $\Lambda$) as to order--$r$ contribution to differential operator, denoting $\Lop (n)_r$, where $r$ is the order of a term as of a differential operator. In the sample term above, the order is $2 b$ or $2 b + 1$ depending on the power of $\Lambda$. We'll see that the specification of order determines the contribution unambiguously. We also exploit a shorthand notation:
\[ \Box_k = \Box + k m^2 \]
For the sake of reference, take the equation for $n=5$, i.e. an operator killing 5--banana:
\[\begin{split}
\Lop(5) =&\  x^4 \Box_{25} \Box_9 \Box_1 + 2 (2 D-2) x^2 \Lambda (\Box_{25} \Box_9 + \Box_{25} \Box_1 + \Box_9 \Box_1) +\\
&\ + (2 D -2) x^2 ((29 D - 36)\Box^2 + (434 D - 532) m^2 \Box + (725 D - 880)m^4)+\\
&\ +4 (2 D - 2)(3 D - 4)(4 D - 6)\Lambda \Box_7 + (2 D -2)(3 D -4)(4 D -6) (5 D -8)\Box_5\\
\end{split}\]
From this expression we have:
\[\begin{split}
\Lop(5)_6 =&\  x^4 \Box_{25} \Box_9 \Box_1 = x^4(\Box + 25 m^2)(\Box + 9 m^2)(\Box + m^2);\\[1mm]
\Lop(5)_5 =&\  2 (2 D-2) x^2 \Lambda (\Box_{25} \Box_9 + \Box_{25} \Box_1 + \Box_9 \Box_1) =\\[1mm]
=&\ 2 (2D - 2) x^2 \Lambda ((\Box + 25 m^2)(\Box + 9 m^2)+(\Box + 25 m^2)(\Box + m^2)+(\Box + 9 m^2)(\Box + m^2));\\[1mm]
\Lop(5)_4 =&\  (2 D -2) x^2 ((29 D - 36)\Box^2 + (434 D - 532) m^2 \Box + (725 D - 880)m^4);\\[1mm]
\Lop(5)_3 =&\  4 (2 D - 2)(3 D - 4)(4 D - 6)\Lambda \Box_7 = 4 (2 D - 2)(3 D - 4)(4 D - 6)\Lambda (\Box + 7 m^2);\\[1mm]
\Lop(5)_2 =&\  (2 D -2)(3 D -4)(4 D -6) (5 D -8)\Box_5 = (2 D -2)(3 D -4)(4 D -6) (5 D -8)(\Box + 5 m^2);\\
\end{split}\]
Now we list the properties of $\Lop(n)$ we have observed.

\begin{itemize}

\item The equations are polynomial in $x^2, \Lambda, \Box, m^2$ and $D$. It is homogeneous in the dimensional variables $x^2,\Box, m^2$, with obvious dimensions
\begin{equation}
	[x^2]=-2,\ [\Box]=[m^2]=2
\end{equation}
The overall degree of all equations is $2$. For this homogeneity, the contribution $\Lop(n)_k$ is homogeneous in $\Box$ and $m^2$ of degree $\bfloor{\frac{k}{2}}$ (for odd $k$, the extra differentiation comes from $\Lambda$), and starts with $x^a$ for $a = 2 \bfloor{\frac{k}{2}} - 2$.

\item The order of $\Lop(n)$ is $n+1$. For example $\Lop(5)$ above is of order $6$.

\item \label{xdeg}The highest degree in $x$ is $x^{n-1}$, if we count $\Lambda=x \partial_x$ as carrying $x$ in $\Lop(n)_{n+1}$ for even $n$. Further we will check this to be consistent with calculations in momentum space. For example, $\Lop(5)$ above starts with $x^4$.

\item $\Lop(n)_k$ is a polynomial of degree $n-k+1$ in $D$. For example, $\Lop(5)_4$ is quadratic in $D$.

\item The leading--order term admits a simple factorized form:
\begin{equation}
	  \mathcal{L}(n)=x^{n-2}\Lambda \prod_{i=1}^{n/2} \Box_{(2i)^2}  +\text{lower--order terms} = x^{n-2} \Lambda \left((\Box + (2 m)^2)(\Box + (4 m)^2)\ldots(\Box + (n m)^2)\right) + \text{lO}
\end{equation}
for even $n$, and
\begin{equation}
	  \mathcal{L}(n) =x^{n-1} \prod_{i=0}^{\bfloor{\frac{n}{2}}} \Box_{(2i+1)^2} + \text{lower--order terms} = x^{n-1} \left((\Box + m^2)(\Box + (3m)^2)\ldots(\Box + (n m)^2)\right) + \text{lO}
\end{equation}
for odd $n$.

For example, above $\Lop(5)_6 = x^4 \Box_{25} \Box_9 \Box_1$, so $\Lop(5) = x^4 \Box_{25} \Box_9 \Box_1 + \text{terms of lower order}$

\item The next--to--leading term cannot factorized in such a way. However, for odd $n$ it can still be cast to a relatively simple form:
  \begin{equation}
    \begin{split}
      \mathcal{L}(n)=&\ \text{leading--order} + \bfloor{\frac{n}{2}}(2D-2) x^{n-3} \Lambda \sum\limits_{j=1}^k \prod_{i\neq j}^n \Box_{(2i+1)^2}  + \text{lower--order} =\\
        =&\ \text{LO} + \bfloor{\frac{n}{2}}(2 D -2) x^{n-3} \Lambda \sum\limits_k \left((\Box+ m^2)(\Box + (3 m)^2)\ldots \xcancel{(\Box + (k m)^2)} \ldots(\Box + (n m)^2)\right) + \text{lO}\\
    \end{split}
  \end{equation}
  For example, above we have $\Lop(5)_5 = 2 (2 D-2) x^2 \Lambda (\Box_{25} \Box_9 + \Box_{25} \Box_1 + \Box_9 \Box_1) $.

  For even $n$, the analogous formula is more complicated. Still, we can express second--to--leading contribution in terms of elementary symmetric polynomials of specific arguments:
\begin{equation} \label{evensubl}
		\mathcal{L}(n) = \text{LO} + (2D-2)x^{n-2} \left( \sum_{j=0}^{n/2} \left(\left(\frac{n}{2}\right)^2-j\frac{n}{2}+\dfrac{j}{2} \right) e_{j}\left\{ x_i=2^2,4^2,\ldots, n^2\right\} (m^2)^j\Box^{\frac{n}{2}-j} \right)  + \text{lO}
\end{equation}
for odd $n$. For example, $\Lop(6)_6=\ldots$. This is an analogue of the above formula for odd $n$, as, by expanding all the brackets, we arrive at similar expression in terms of elementary symmetric polynomials:
\[ \Lop(n)_n = x^{n-3} \Lambda \sum\limits_{j=0}^{\bfloor{n/2}} (\bfloor{\frac{n}{2}} - 1 + j) e_j \{x_i = 1^2, 3^2,\ldots,n^2\} (m^2)^j \Box^{\bfloor{\frac{n}{2}}-j}\]
In short (and well--suited for momentum representation) form, these formulas can be summarized as:
\begin{equation}
	\Lop(n)_n = \bfloor{\frac{n}{2}} (2 D -2) x^{n-3} \Lambda \dfrac{d}{d \Box} \prod_{i=0}^{\bfloor{n/2}} \Box_{(2i+1)^2}
\end{equation}
for odd $n$, and
\begin{equation}
	\Lop(n)_n = (2 D -2) x^{n-2} \left(\frac{1}{2}\frac{n}{2}\prod_{i=1}^{n/2} \Box_{(2i)^2} + \frac{n+1}{2}\Box\frac{d}{d \Box} \prod_{i=1}^{n/2} \Box_{(2i)^2}\right)
\end{equation}
for even $n$.
\item Terms on the other end of the equations have an even simpler structure. For the last and preceding term we have:
  \begin{equation}
    \begin{split}
      \mathcal{L}(n)=&\ \text{higher--order terms} + (n-1)\left(\prod_{j=2}^{n-1} (jD-2j+2)\right) \Lambda \Box_{n+2} + \left(\prod_{j=2}^n (jD-2j+2)\right) \Box_n =\\
      =&\ \text{HO}+(n-1)\Big((2D-2)(3D-4)(4D-6)\ldots((n-1)D-2 n +4)\Big) \Lambda \Box_{n+2} +
      \\
      &\ +\Big((2D-2)(3D-4)\ldots(n D - 2 n +2)\Big) \Box_n
      \end{split}
\end{equation}
  so that

  \[\label{last} \Lop(n)_2 = \left(\prod_{j=2}^n (jD-2j+2)\right) \Box_n \]
  \[\label{quasilast} \Lop(n)_3 = (n-1)\left(\prod_{j=2}^{n-1} (jD-2j+2)\right) \Lambda \Box_{n+2} \]

For example, $\Lop(5)_2 = (2 D -2)(3 D -4)(4 D -6) (5 D -8)\Box_5 $ and $\Lop(5)_3 = 4 (2 D - 2)(3 D - 4)(4 D - 6)\Lambda \Box_7$, which means $\Lop(5)$ has the form $\text{higher--order}+4 (2 D - 2)(3 D - 4)(4 D - 6)\Lambda \Box_7+ (2 D -2)(3 D -4)(4 D -6) (5 D -8)\Box_5$

\item We can dig even further in the structure of the coefficients. To give an example $\ldots$

\end{itemize}

\subsection{Proofs}

Here we provide some combinatorial calculations that produce the terms in the subsection above. We'll base them directly on the recursion relations derived above. The conceptual, rather then computational, reasoning for these patterns remain, however, unknown to us.

Our reasoning is based on how the contributions of the form $ x^a \Lambda^{0 | 1} (c_b \Box^b + c_{b-1} m^2 \Box^{b-1} + \ldots) $ behave under the recursion
\[\label{termrec1} L_{k+1}^{(n)} = \Lambda L_k^{(n)} + k (D-2) L_{k-1}^{(n)} + x^2 k (n - k + 1) m^2 L_{k-1}^{(n)} \]
First, we deal with contribution of $L_k$ to $L_{k+1}$. Expanding the first and the second term, we obtain:
\[\label{termrec2} x^a (\ldots) \mapsto \Lambda x^a (\ldots) + k (D-2) x^a (\ldots) = x^a \Lambda (\ldots) + (k (D-2) + a) x^a (\ldots) \]
\[ x^a \Lambda (\ldots) \mapsto \Lambda x^a \Lambda (\ldots) + k (D-2) x^a \Lambda (\ldots) =  x^{a+2} \Box (\ldots) + (a - (D-2) + k(D-2)) x^a \Lambda (\ldots)\]
where we used the operator identities $\Lambda x^a = x^a (\Lambda + a)$ and $\Lambda^2 = x^2 \Box - (D-2) \Lambda$. Notice that, in this subsection, by $(\ldots)$ a dimensionally correct polynomial in $\Box$, $m^2$ and $D$ is denoted.

Second, we deal with contribution of $L^{(n)}_{k-1}$. It has even simpler structure than considered above:
\[\label{termrec3} x^a (\ldots) \mapsto x^{a+2} k (n - k + 1) m^2 (\ldots)\]
\[\label{termrec4} x^a \Lambda (\ldots) \mapsto x^{a+2} k (n - k + 1) \Lambda m^2 (\ldots) \]
Below we present some conclusions from inspecting these relations.

\subsubsection{$D$ degree property}

First, we show that order--$r$ contribution to $\Lop(n)$ has, as a polynomial in $D$, a degree $n - r +1$. This is a particular case of the fact that order--$r$ term (defined by the same conventions \eqref{order}) to $L_{k+1}^{(n)}$ is of a degree $k-r$, for $k = n+1$. To show this, we divide $L_{k}^{(n)}$ into individual terms and notice from inspection of the relations above that order--$r$ term in it contributes either to order--$r+1$ term in $L_{k+1}^{(n)}$ or to order--$r$ term with $D$ power increased by $1$, but not both: namely

\begin{equation*} x^a (\ldots) \mapsto \overbrace{x^a \Lambda (\ldots)}^{\text{order increased}} + \overbrace{(k(D-2) + a) x^a (\ldots)}^{\text{$D$ degree increased}} \end{equation*}

\begin{equation*} x^a \Lambda (\ldots) \mapsto \overbrace{x^{a+2} \Box (\ldots)}^{\text{order increased}} + \overbrace{(a - (D-2) + k(D-2)) x^a \Lambda (\ldots)}^{\text{$D$ degree increased}}\end{equation*}

This means that if $D$ degree property holds for $L_k^{(n)}$, then it holds for $(\Lambda  + k (D-2)) L_k^{(n)}$, which is a $L_k^{(n)}$ contribution into  $L_{k+1}^{(n)}$ as well. For $L^{(n)}_{k-1}$ contribution we notice that order--$r$ term in it contributes only to order--$r+2$ term in $L_{k+1}^{(n)}$, so this part of the recursion preserves $D$ degree property too: if order--$r$ term in $L_{k-1}^{(n)}$ had degree $k-1-r$ in $D$, then its contribution to $L_{k+1}^{(n)}$ is to be of the same degree, but $k+1 - (r+2) = k -1 -r $, so the $D$ degree property holds again.

The two remaining things to take care of is the possibility of $D$ degree drop-down from the cancellation of some terms above and the base step of our inductive argument. About the first, we notice that as the contribution of order--$r$ term of $L_{k-1}^{(n)}$ to $L_{k+1}^{(n)}$ is of lower degree in $\Box$ than the contribution of order--$r+2$ term of $L_{k}^{(n)}$, it cannot cancel the coefficient before highest $D$ power (notice that the coefficients before $D$ powers are polynomials in $\Box$). For the second, we notice that $L^{(n)}_0 = 1$ is of degree $0$ in $D$ and contains order--$0$ term only, while $L^{(n)}_1 = \Lambda$ is also of degree $0$ and contains degree--$1$ term only, so the $D$ degree formula holds for them.

\subsubsection{Lowest--order terms}

Pushing to extreme the reasoning above allows us to compute the coefficient before the last term in $\Lop(n)$ \eqref{last}. To do this we notice that recursion relations \eqref{termrec1}--\eqref{termrec4} either increase the order of a term or preserve it, but not decrease; as the last term is evidently of the form $\langle\text{coefficient}\rangle x^2 (\Box + c m^2)$, we need to keep track only of the order--preserving terms starting from the first $L_k^{(n)}$ containing such a contribution for the first time, namely, from $k=2$:
\[ L_2^{(n)} = x^2 (\Box + n m^2) \]
As the order--preserving part in \eqref{termrec1} is $ x^a (\ldots) \mapsto (a + k (D-2))x^a (\ldots)$ and the $L_{k-1}^{(n)}$ contributions to $L_k^{(n)}$ \eqref{termrec3}--\eqref{termrec4} are always order--increasing, this term gets successively multiplied by $2 + 2(D-2)$, $2 + 3(D -2)$,\ldots,$2 + n (D-2)$, therefore,
\begin{equation}
	L^{(n)}_{n+1}= \text{higher--order} +  \left(\prod_{k=2}^{n} ( k(D-2)+2) \right) \tilde{L}^{(n)}_2 = \text{HO} + x^2 \left(\prod_{k=2}^{n}  (k(D-2)+2) \right)  \left(\Box+n m^2 \right)
\end{equation}
which nothing but \eqref{last}. Notice that as this is the lowest--order contribution, the $x^2$ multiplier can be canceled in the whole operator without introducing the $x^{-2}$ coefficients: $\Lop(n) = \frac{1}{x^2}L^{(n)}_{n+1}$; however, if we wanted to avoid this effect, we would have to introduce $x^{-2}$ into the initial conditions of the recursion.

This argument can be suitably modified to compute the second--to--last term \eqref{quasilast}. It evidently has the structure $\langle\text{coefficient}\rangle x^2 \Lambda (\Box + c m^2)$. Therefore, at each recursion step it gets a contribution from order--preserving terms of \eqref{termrec2} or order--increasing terms of \eqref{termrec1}, but not from \eqref{termrec3}--\eqref{termrec4}, as, starting from $k=2$, each term in $L_k^{(n)}$ is at least order--$2$ in $x$. Therefore, we need to track $x^2 \Lambda (\ldots)$--term starting from the first $L_k^{(n)}$ containing such a term, namely, from $k=3$:
\[ L_3^{(n)} = x^2 \Lambda (\Box + (3n-2)m^2) + (2 D - 2) x^2 (\Box + n m^2) \]
At each step of the recursion the order--$3$ term of $L_{k}^{(n)}$ , first, gets multiplied by $2 - (D-2) + k (D-2)) = (k-1)(D-2)+2$, and, second, gets a correction from the last term; however, for the discussion above, the last term of $L_k^{(n)}$ is given by
\[ L_{k}^{(n)} = \text{higher--order} + \prod_{l=2}^{k-1} (2 + l (D-2)) x^2 \Box_n = \text{HO} + (2 D -2)\ldots((k-1)(D-2)+2) x^2 \Box_n \, .  \]
Therefore the correction is of the form
\[ (2D-2)\ldots((k-1)(D-2)+2) x^2 \Lambda \Box_n \]
Consequently, we have
\[\begin{split}
 L_4^{(n)} = &\ \text{higher--order} + (2 + 2(D-2))x^2 \Lambda \Box_{3n-2} + (2 D - 2) x^2 \Lambda \Box_n + \text{last} =\\
=&\ \text{HO} + (2 D - 2) x^2 \Lambda (\Box_{3n -2} + \Box_n) + \text{last}\ ,
\\
\\
L_5^{(n)} =&\ \text{HO} + (2D-2)(2 + 3 (D-2)) x^2 \Lambda (\Box_{3n-2} + \Box_n) + (2D-2)(3D-4) x^2 \Lambda \Box_n + \text{last} =\\
	=&\ \text{HO} + (2 D -2)(3 D-4) x^2 \Lambda (\Box_{3n -2} + 2 \Box_n) + \text{last}\ ,
\\
\\
 L_6^{(n)} =&\ \text{HO} + (2D-2)(3D-4)(2 + 4(D-2)) x^2 \Lambda (\Box_{3n-2} + 2 \Box_n) + (2D-2)(3D-4)(4D - 6) x^2 \Lambda \Box_n + \text{last} =\\=&\
\text{HO} + (2D-2)(3D-4)(4D-6) x^2 \Lambda (\Box_{3n -2} + 3 \Box_n) + \text{last}\ ,
\end{split}\]
so we observe that the structure
\[L_k^{(n)} = \text{HO} + ((2D-2)\ldots((k-2)(D-2)+2)) x^2 \Lambda (\Box_{3 n-2} + (k-3)\Box_n) + \text{last}\]
is preserved by recursion relations and continues up to
\[ L_{n+1}^{(n)} = \text{HO} + (2D-2)(3D-4)\ldots((n-1) (D-2) + 2)) x^2 \Lambda (\Box_{3n-2}+(n+1-3)\Box_n) + \text{last} \]
But as $\Box_{3n-2} + (n-2) \Box_n = (n-1)\Box + (3n -2)m^2 + (n-2) n m^2 = (n-1) (\Box + (n+2) m^2)$, the order--$3$ term has the form
\[ \prod\limits_{k=2}^{n-1} (k(D-2)+2) (n-1) x^2 \Lambda (\Box + (n+2) m^2) \]
which is nothing but \eqref{quasilast}

\subsubsection{Leading--order terms} \label{leadproof}

For the leading terms, a combinatoric derivation from the recursion relations is complicated. However,by $D$ degree property, the leading--order and next--to--leading order terms have degrees $0$ and $1$ in $D$, respectively. Therefore, knowing the leading term in some fixed dimension and subleading one in two fixed dimensions is enough to fixed their form completely. For the reason, we can use our results for $D=1$ and $D=3$, where a complete answer for arbitrary $n$ is available. Multiplying the $D=1$ answers \eqref{D1odd}--\eqref{D1even} by $x^{n-1}$ or $x^{n-2}$ for higher--$D$ compatibility, we obtain
\[ \left.\Lop(n)\right|_{D=1} = x^{n-1} \prod\limits_{k=0}^{\bfloor{\frac{n}{2}}} (\Box + (n-2 k)^2 m^2) \]
for odd $n$, and
\[ \left.\Lop(n)\right|_{D=1} = x^{n-2}\Lambda \prod\limits_{k=0}^{\frac{n}{2} - 1} (\Box + (n - 2 k)^2 m^2)\]
As these operators have the form \eqref{order}, they present the leading--order terms in $\Lop(n)$, and as such are $D$--independent. Therefore for \emph{arbitrary} $D$ we have
\[ \Lop(n)_{n+1} = x^{n-1} \prod\limits_{k=0}^{\bfloor{\frac{n}{2}}} (\Box + (n-2 k)^2 m^2),\ n\text{ odd}\]
\[ \Lop(n)_{n+1} = x^{n-2}\Lambda \prod\limits_{k=0}^{\frac{n}{2} - 1} (\Box + (n - 2 k)^2 m^2),\ n\text{ even}\]
as well.

To compute the subleading term for odd $n$, it is sufficient to extract it from the full answer for $D=3$, where we have:
\begin{equation}
	\begin{split}
		&\ \prod_{i=0}^{k} \left( \partial^2+ (2i+1)^2 m^2 \right) \left( x^{2k+1} G_{2k+1} \right) = \prod_{i=0}^{k} \left( \Box -\dfrac{2}{x^2} \Lambda + (2i+1)^2 m^2 \right) \left( x^{2k+1} G_{2k+1} \right) = \\
		&\ = x^{2k+1} \prod_{i=0}^{k} \left( \Box  + (2i+1)^2 m^2 \right)  G_{2k+1} - 2 x^{2k-1} \Lambda \dfrac{d}{d \Box} \left( \prod_{i=0}^{k} \left( \Box  + (2i+1)^2 m^2 \right) \right) + \\
		&\ + 2 (2k+1) \Lambda \dfrac{d}{d \Box} \left( \prod_{i=0}^{k} \left( \Box  + (2i+1)^2 m^2 \right) \right) +\text{lower--order} =\\
		&\ =x^{2k+1} \prod_{i=0}^{k} \left( \Box  + (2i+1)^2 m^2 \right)  G_{2k+1} - 4 k x^{2k-1} \Lambda \dfrac{d}{d \Box} \left( \prod_{i=0}^{k} \left( \Box  + (2i+1)^2 m^2 \right) \right) +\text{lo}
	\end{split}
\end{equation}
where in the second line we expand brackets such that we take the  $\dfrac{2}{x^2}\Lambda$ term  and used the commutator:

\begin{eqnarray}
	\Box x^{2k+1} = x^{2k+1} \Box  + 2(2k+1) x^{2k-1} \Lambda +(2k+1)(2k+2)x^{2k-1}.
\end{eqnarray}

As in $D=1$ we have $\left. \Lop(n)_n \right|_{D=1} =0$ and the term in consideration is linear in $D$, it has a form $a (D-1)$. But as at $D=3$ $2D-2=4$, we just need to replace the $4$ coefficient above with $2D-2$ to obtain an arbitrary odd $n$ answer:
\[ \Lop(n)_n = (2 D -2) \bfloor{\frac{n}{2}} x^{n-3} \Lambda \dfrac{d}{d \Box} \left( \prod_{i=0}^{k} \left( \Box  + (2i+1)^2 m^2 \right) \right) \]
(one extra $x$ power has been canceled for the sake of uniformity)

In an analogous way, the formula for even--$n$ subleading term \eqref{evensubl} can be derived.

\section{Generalization: different--mass case}\label{sec:DifferentMass}
It is easy to see that the general reasoning in \ref{demagogy} doesn't rely heavily on the equality of masses of different virtual particles. Therefore our procedure should admit a generalization to diagrams with unequal internal masses. We argue this is indeed the case by demonstrating an explicit generalization of our basic examples, namely, $D=1$, and $n=2$. We also restate our generic arguments in a form of a routine for deriving a differential equation for arbitrary $D$ and generic masses. However, as the equations arising as results are significantly more complicated than equal--mass ones, only very basic observations about the order and leading--order term are presented.

\subsection{$D=1$ case, any $n$}
Here we present a generalization of analysis in \ref{ }, leading to a derivation of arbitrary--$n$ equation in $D=1$. The base equations are
\begin{equation}
	(\partial^2+m_k^2)G(m_k;x) = 0
\end{equation}
so off--singularity Green's functions are given by exponents $G(m_k;x) = e^{\pm i m_k x }$ with arbitrary mass tuple $m_k$. The banana function is given by
\begin{equation}
	G_n(\vec{m};x)=\prod_{k=0}^{n} G(m_k;,x)
\end{equation}
So for $\Lop$ a differential operator killing the banana function we should have
\begin{equation}
	\mathcal{L}_n(\vec{m})  e^{\left( i  \left( \pm m_1 \pm m_2 \pm \ldots \pm m_n \right) x \right)} =0
\end{equation}
for arbitrary combination of $+$'s and $-$'s. Such $\Lop$ is given by a straightforward generalization of \eqref{D1eqm}:
\begin{equation}\label{D1diffm}
	\prod_{\vec{\epsilon}} \left( \partial + i \left( \epsilon_i m_i \right) \right) G_n(\vec{m};x) =0
\end{equation}
where the sum goes over all $2^n$ different sign--vectors $\Vec{\epsilon} \in \left\{ -1,1 \right\}^{n}$. Laplacian--like form of this equation is
\begin{equation}\label{D1diffm}
	\prod_{\vec{\epsilon}} \left( \partial^2 + \left( \epsilon_i m_i \right)^2 \right) G_n(\vec{m};x) =0
\end{equation}
with $\vec{\epsilon} \in \{-1, 1\}^{n} / {\pm}$.

Notice that the order of this differential operator is $2^n$ as opposed to $n+1$ in the equal mass case, as expected from analysis in sec. \ref{demagogy}. This means that in the equal mass case the space of solutions gets degenerated. For example, $e^{i (m_1 -m_2) x}$ and  $e^{i (m_2 -m_1) x}$ both become the constant solution at $m_1=m_2$. This means that in the equation we get a degeneration of operators
\begin{equation}
	(\partial+i (m_1-m_2))	(\partial-i (m_1-m_2)) \,  \overset{m_1=m_2}{\longrightarrow} \,   \partial_x
\end{equation}
For example, for $D=1$, $n=2$ we have:
\begin{equation}
	\left(\Box+(m_1-m_2)^2 \right)\left(\Box+(m_1+m_2)^2 \right) G_2(m_1,m_2;x) = 0 \,  \longrightarrow \, \Lambda \left(\Box+4m^2 \right) G_2(x) = 0
\end{equation}

\subsection{$n=2$ case, any $D$}
Here we present a derivation of $n=2$ general--$D$ equation analogous to example in \ref{sectionGeneralD}.
We are analyzing 2--banana function:
\begin{equation}
	G_2(m_1,m_2;x) = G(m_1;x) G(m_2;x)
\end{equation}
To derive the differential equation we will use the same recursive approach as in sec.\ref{sectionGeneralD}. We employ, however, radial--coordinate calculations. As in \eqref{n2diff} we get:
\begin{equation} \label{boxdifm}
	\left(\Box+(m_1^2+m_2^2) \right) G_2(m_1,m_2;x)  =  2 \partial_\mu G(m_1;x) \partial_\mu G_(m_2;x) =  2 \partial G(m_1;x) \partial G(m_2;x)
\end{equation}
Further differentiation leads to the following relation
\begin{eqnarray}
	\Box \left( \d G_{m_1} \d G_{m_2} \right) =  -\left(\frac{2(D-1)(D-2)}{x^2}+m_1^2+m_2^2 \right) (\d G_{m_1} \d G_{m_2}) - \frac{2(D-1)}{x}\partial G_{m_1} \partial G_{m_2} + 2 m_1^2 m_2^2 G_{m_1} G_{m_2}
\end{eqnarray}
Expressing $\d G_{m_1} \d G_{m_2}$ from \eqref{boxdifm} and supplying it here we come out with:
\begin{equation}
	\left\{x^2(\Box +m_1^2+m_2^2 )+2(D-1)\Lambda + 2(D-1)(D-2)\right\}(\Box + m_1^2+m_2^2) G_{m_1} G_{m_2}
	= 4m_1^2m_2^2 x^2 G_{m_1} G_{m_2}
\end{equation}
Thus, restated in a more furnished form,
\begin{equation}
	\boxed{
	\begin{array}{c}
		\left\{x^2\Big(\Box + (m_1+m_2)^2\Big)\Big(\Box + (m_1-m_2)^2\Big)
		+ \right.\\ \left.
		+ 2(D-1)\Lambda\Big(\Box + m_1^2+m_2^2\Big) + 2(D-1)(D-2)\Big(\Box + m_1^2+m_2^2\Big)
		\right\}G_2 = 0
	\end{array}
}
\label{G2eqDdm}
\end{equation}
Clearly, at $D=1$ this equation reduces to \eqref{D1diffm}. Some mass terms seem to reduce smoothly to their uniform--mass case limits: for example, in the leading terms $ (m_1+m_2)^2 \rightarrow 4m^2 $, while in the last term $ m_1^2+m_2^2 \rightarrow  2m^2 $. Generally, however, the equal--mass limit does not appear to be a trivial one. Instead of ``extra'' terms just vanishing, the operator gets additional \emph{factorization} at equal masses. For example, in our case we have:
\begin{eqnarray}
	\begin{split}
		0&\ =x^2(\Box + 4m^2)\Box G_2 +2(D-1)\Lambda(\Box +2m^2)G_2  + 2(D-1)(D-2)(\Box+2m^2) G^2=\\
		&\ =\Big(\Lambda+(D-2)\Big)\Big(\Lambda (\Box+4m^2)G_2 + 2(D-1)(\Box + 2m^2) G_2\Big) 		
	\end{split}
\end{eqnarray}

\subsection{The general procedure}
The general procedure is a specification, for the case of interest, of general considerations in \ref{demagogy}. It can be formulated as follows: suppose we are looking for a differential equation for $n$--banana.
\begin{enumerate}
\item denote
  \[
  I_{s_1, s_2, \ldots, s_k} = G_{1} G_{2} \ldots (\d G_{s_1}) (\d G_{s_2}) \ldots (\d G_{s_k})
  \]
  where $G_i := G(m_i;x)$, $\d = \d / \d |x|$ and $s_1,\ldots s_k$ is an arbitrary subset of $1, 2, \ldots, n$

\item These functions are differentially related, namely,
  \[
  \d I_{s_1, s_2, \ldots s_k} = \sum\limits_{a \ne s_1, \ldots s_k} I_{a,s_1,\ldots,s_k} + k \frac{D-1}{x} I_{s_1,\ldots,s_k} - \sum\limits_{j=1}^k m_{s_j}^2 I_{s_1, s_2, \ldots, \xcancel{s_j} \ldots s_k}
  \]
\item
  By taking derivatives of these relations and substituting them in place of derivatives, one can express any derivative $\d^k I_{\emptyset} $ of banana function as a linear combination of various $I_{\bullet; \ldots; \bullet}$'s:
  \[
  \d^k I_{\emptyset} = \sum\limits_{\mathbf{J} \subseteq \{1,2,\ldots n\}} \alpha_\mathbf{J}^{(k)} I_\mathbf{J}
  \]
  The first $2^n - 1$ derivatives, including the zeroth are then sufficient to solve this linear system and express arbitrary $I_J$ backwards as a linear combination of $\d^k I_{\emptyset}$:
  \[
  I_{\mathbf{J}} = \sum\limits_{k = 0}^{2^n-1} \beta_{k}^{(\mathbf{J})} (\d^k I_{\emptyset})
  \]
\item
  The $2^n$'th derivative expression therefore becomes a differential equation, when $I_{\bullet; \ldots; \bullet}$ in form of linear combinations of lower derivatives are substituted into it:
  \[
  \d^{2^n} I_{\emptyset} = \sum\limits_{\mathbf{J} \subseteq \{1,2,\ldots n\}} \alpha_\mathbf{J}^{(2^n)} I_\mathbf{J}\ \rightarrow\ \d^{2^n} I_{\emptyset} =  \sum\limits_{k = 0}^{2^n-1} \gamma_{k} (\d^k I_{\emptyset})
  \]
  As $I_{\emptyset}$ is the initial banana function, this is the differential equation we settled for. Alternatively, one can use the identity
  \[
  \d I_{1,2,\ldots,n} = n \frac{D-1}{x} I_{1,2,\ldots,n} - \sum\limits_{k=1}^n m_k^2 I_{1,2,\ldots,\xcancel{k},\ldots,n}
  \]
  which does not contain higher derivatives.
\end{enumerate}

\section{Translation to momentum space}\label{sec:Tomomentum}
In order to compare our position space equations with some well known results we have to translate those into momentum space. Since all of our equations are written in  terms of scalar operators $\Box, \Lambda$ and $x^2$ it is a straightforward calculation. Also, since we are dealing with the Feynman propagator we may assume, that momentum space integrals only depend on the square of the external momentum. Denoting it as $t=p^2$ we then have the following simple rules:
\begin{equation}\label{xtop}
	\begin{split}
	\Box \cdot F(x) & \rightarrow -t \cdot F(t)
	\\
	\Lambda \cdot F(x) & = -\left(D+2t \dfrac{d}{dt} \right) \cdot F(t) = -\left((D-2)+2\dfrac{d}{dt} t\right) \cdot F(t)\\
	x^2 \cdot F(x) & \rightarrow -\left( 2D \dfrac{d}{dt}+4 t \dfrac{d^2}{dt^2} \right) F(t)
	= -2\frac{d}{dt}\left( (D-4)  +2 \dfrac{d }{dt } t \right) \cdot F(t)
	\end{split}
\end{equation}
where $F(t)$ is the  Fourier transform of $F(x)$. These transition rules for operators follow from the basic properties:
\begin{eqnarray}
	\begin{split}
		x^\mu F(x) &=   \int d^Dp \, e^{i p x} \left(-\frac{\partial }{\partial p^\mu}  \right)F(p)\\
		\frac{\partial }{\partial x^\mu} F(x) &= \int d^Dp \, e^{i p x}  p^\mu F(p)
	\end{split}
\end{eqnarray}
After transitioning to momentum space one would have to reorder the derivative to bring the equation into a canonical form:
\begin{equation}
	\hat{\mathcal{L}}(n)G_n(t) = \left( \sum_{k=0}^{\operatorname{ord}(n)} q_{k}(t) \dfrac{d^k}{dt^k} \right) G_n(t)
\end{equation}
where the functions $q_k(t)$ are express trough the coefficients of the position space operator in an intertwined way.
\\

The order of differential operators in momentum and position space is different. This would make us generally suspicions, since the order counts the number of independent solutions. However even on in the simplest case of the equation for the cut propagator we see this discrepancy:
\begin{equation}
	(\Box+m^2  ) G(x) = 0 \rightarrow (p^2-m^2)G(p)=0
\end{equation}
the two independent solutions for the momentum space equation are given by singular delta-functions \eqref{G+-}, therefore the counting is affected. 
\\\\
It is also well-known that banana integrals in momentum space may acquire UV divergences. In terms of dimensional regularization this amounts to the appearance of poles at some integer dimensions. Since, for the most part, we do not deal with UV behavior, we also will discuss  momentum space divergences elsewhere. As we have mentioned the advantage of working in position space is  the absence of these divergences. This is because in momentum space we deal with integrals instead of just functions, which can be divergent in large enough number of dimensions by power counting.
\subsection{Comparison with current literature}
In \ref{sec:Observations} we made several statements about the coefficients in the position space operators. Now we would like to derive some their counterparts in momentum space, considering $q_k(t)$ and $\operatorname{ord}(n)$. Comparing the results with \cite{Lairez:2022zkj} we find complete agreement.
\\\\
Formulas \eqref{xtop} mean that the term of the highest degree in $t$ derivatives comes from the highest degree terms in $x$. Multiplication by $x^2$ continuities a second derivative in $t$, while $\Lambda=x\partial_x$ produces a first derivative. Therefore, according to sec. \ref{sec:Observations}:
\begin{equation}
	\operatorname{ord}(n)=\deg_{x}
	\left\{ \mathcal{L}(n)  \right\}= n-1
\end{equation}
We can compute coefficient of the highest order term. This does not require complicated operator reordering, since the highest derivative comes from simply moving all derivatives to the right. Therefore we have:
\begin{equation}
	q_{n-1}(t)=	-(-2)^{n-1} t^{\frac{-1+(-1)^n}{2}+\left\lfloor \frac{n+1}{2} \right\rfloor  } \prod_{i=0}^{\left\lfloor \frac{n-1}{2} \right\rfloor } \left(   t - (n-2i)^2 m^2 \right)
\end{equation}
The succeeding term can also be calculated. It gets two contributions. First is from lower order terms in $x$ in position space, for which we pull the $t$-derivative to the right. The second contribution comes from the highest order term in position space, but taking the one order  lower derivative term, that appears after reordering of derivatives.
\begin{equation}
	q_{n-2}(t)=	\frac{(n-1)}{2} \cdot \left( \dfrac{n(D-2)}{2t}q_{n-1}(t) - (D-3) \dfrac{d q_{n-1}(t)}{dt}\right) 
\end{equation}
which at $d=2$ coincides with \cite{Lairez:2022zkj} 
The lower derivative terms in momentum space require increasingly more reordering and get contributions from lower terms in $x^2$ in position space. Instead, just as in position space, we can switch our attention to the lowest order term. Hence for the contribution without derivatives we obtain:
\begin{equation}
	q_0(t)=- \left( \prod_{j=2}^{n-1} (j D-2j-2) \right) \cdot  \left( \left(nD-2D+4\right)m^2+(D-4)t \right)
\end{equation}
which reduces to $-(-2)^{n-1} (t-n m^2)$ at $D=2$,  which agrees with \cite{Lairez:2022zkj}.\\\\
Hence the position space calculations are quite efficient to also obtain general answers for momentum space operators.

\section{Conclusion}

In this text we were concerned with the hidden structures of perturbative quantum field theory manifested in properties of Feynman diagrammatic integrals. Under the hood these properties represent the shades of integrability of underlying quantum field theory\cite{morozov1992string,Gerasimov:2000pr}. They seemingly have deep origins in the structure of stringy amplitudes, especially
to their the universal moduli space formulation \cite{Morozov:1987pk,D_Friedan_1987},
resembling provokingly the practice of motivic theory \cite{Brown:2015fyf,Morozov:2009kc,Doran:2023yzu}. In a sense, the hidden symmetries we refer to are particle limits of the corresponding string structures. These deep arguments are bridged to diagrammatics by the algebraic structures on graphs
\cite{Kreimer:1997dp,Gerasimov:2000pr},
which are usually studied in the theory of matrix models
and their integrability properties \cite{morozov1994integrability}. Here our approach, however, is different: proceed down--top and explore the properties in question on the level of Feynman integration directly. This work is to provide us with intuition, computation routines and a bunch of examples all at the same time.

In the present paper our main contribution is a position--space--based method of computing the differential equations on $n$--banana Feynman diagrams.
The core idea is to use some very general properties of $D$--modules, namely the differential equations on functions induce differential equation on product --- a fact otherwise known as flatness of tensor product of flat connections. 
We should  note that the while the banana graphs are somewhat exotic for the realistic branches of quantum field theory, they take important place on their own in some special models. Namely, in the so--called Aristotelian tensor models the color structure of vertices forbids any diagrams but bananas \cite{Itoyama:2017xid,Itoyama:2017emp} --- and therefore in a model of Aristotelian tensor \emph{fields} the graphs we discuss constitute the whole of perturbation theory. 

The examples computed allow experimental search for signs of hidden structures behind these equations. Our attempts in this direction led to general formulas for some coefficients in the equations. We proved the leading--order terms have simple factorized form, that can be deduced from 
in $D=1$. Other terms appear as a deformation of $D=1$ result and also exhibit factorization patterns described in section \ref{sec:Observations}.

On the other hand, both the computation of the right--hand side and comparison with momentum--space results turn out to be tricky and related to subtleties of multidimensional distribution theory and complex--analytic Fourier analysis. In particular, the minimal order of polynomial differential equations in position space naively does not coincide with momentum--space one, despite the solution space dimension should be the same. These complications need to be resolved before a general theory is set up. On the other hand, even in such a general theory one needs to keep nonsingular--patch phenomena separated for the sake of clarity.

The general nature of the ideas underlying our computations leads us to believe that these method can be extended beyond banana graphs. We will provide more applications in future publications. In particular, we presume the following observations to remain true (rendered appropriately) in general:
\begin{itemize}
\item the utility of switching to position space (where the key building block is the equation of motion, which has a transparent origin)

\item the separation of non--singular patch affairs from the structure of discontinuities  (e.g. absence of divergences for diagrams without internal vertices)

\item the complete tractability of $D=1$ case (trigonometric Green functions)

\item the deformational nature of general--$D$ results (with $D=1$ as a starting point). In particular, the $D$ dependence of the results in a right form should be polynomial or at least analytic.
  
\item the special status of odd--dimensional Green function (polynomial--trigonometric as opposed to Bessel--like)
\end{itemize}
An example of question suggested by the present study is the following. Take any Feynman diagram and consider a family of alike graphs, in which one selected edge gets progressively "bananized". Do differential equations on these diagrams form a structurized tower analogous to one dealt with in this text? In particular: can these equations be computed in some systematic way, given an equation on the initial diagram?

Summing up, the spacetime considerations of Feynman diagrammatic integrals simplify greatly their exploration by means of differential equations. It makes possible to perform massive computations necessary for empirical studies of these functions. Not only that, it allows a different perspective on their analytic structure and in a way completes the emerging general picture with a reciprocal--space counterpart. This convinces us this line of thought is promising and worth pursuing further research.

\paragraph{Acknowledgments}
This work is supported by the grant of the Foundation for the Advancement of Theoretical Physics
“BASIS” and by the joint grant 21-51-46010-ST-a.

\bibliographystyle{unsrt}
\bibliography{bananabrief}

\newpage
\appendix
\section{Selected list of equations}\label{Appendix}

\subsection{$D=1$, arbitrary $n$, equal masses}
\begin{itemize}
\item $n$ is odd:
  \begin{equation*} x^{n-1} \prod\limits_{k=0}^{\bfloor{\frac{n}{2}}} (\Box + (n-2 k)^2 m^2) G_n =  i \frac{n!}{(2 m)^{n-1}} \delta(x) \end{equation*}

\item $n$ is even:
  \begin{equation*} x^{n-2} \Lambda \prod\limits_{k=0}^{\frac{n}{2} - 1} (\Box + (n - 2 k)^2 m^2) G_n =  i \frac{n!}{(2 m)^{n-1}} \delta(x) \end{equation*}
\end{itemize}

\subsection{Arbitrary $D$, equal masses, no singular term}
{\footnotesize
\begin{itemize}
\item $n=1$:
  \begin{equation*} (\Box + m^2) G_1 = 0 \end{equation*}

\item $n=2$:
  \begin{equation*} \Lambda (\Box+4m^2) G_2 + 2(D-1)(\Box + 2m^2) G_2 = 0\end{equation*}

\item $n=3$:
  \begin{equation*} x^2 (\Box+9m^2) (\Box+m^2) G_3 + 2 (2D-2)\Lambda (\Box+5m^2) G_3  + (2D-2) (3D-4) (\Box +3m^2)G_3 = 0 \end{equation*}

\item $n=4$:
  \begin{equation*}\begin{split} x^2\Lambda(\Box+16m^2)(\Box+4m^2)G_4 + 2 x^2 (2D-2)(2\Box^2+25m^2\Box+32m^4)G_4 +\\
      3 (2D-2)(3D-4) \Lambda (\Box+6m^2) G_4 + (2D-2)(3D-4)(4D-6)(\Box+4m^2)G_4 = 0\\
    \end{split}\end{equation*}

\item $n=5$:
  \begin{equation*}\begin{split} x^4(\Box+25m^2)(\Box+9m^2)(\Box+m^2)G_5 + 2(2D-2)x^2\Lambda(3\Box^2+70m^2\Box+259m^4)G_5+\\
      x^2(2D-2)((29D-36)\Box^2+2(217D-266)m^2\Box+5(145D-176)m^4)G_5+\\
      4(2D-2)(3D-4)(4D-6)\Lambda(\Box+7m^2)G^5 + (2D-2)(3D-4)(4D-6)(5D-8)(\Box+5m^2)G_5 = 0\\
  \end{split}\end{equation*}
  
\item $n=6$:
  \begin{equation*}\begin{split} & x^4 \Lambda \left(\Box+4 m^2\right) \left(\Box+16 m^2\right) \left(\Box+36 m^2\right) G_6 + x^4 \left(9 \Box^3+364 m^2 \Box^2 +3136 m^4 \Box + 3456 m^6\right) G_6\\
     & + (2D-2) x^2 \Lambda \left((59 D -72) \Box^2 +112 (14 D -17) m^2 \Box +8 (823 D-999) m^4\right) G_6 +\\
     & 2 (2D-2) (3 D -4) x^2 \left((37 D -48) \Box^2 +28 (23 D -29) m^2 \Box +12 (111 D -137) m^4\right) G_6 +\\
     & 5 (2D -2) (3 D -4) (4 D - 6) (5 D   -8) \Lambda \left(\Box+8 m^2\right) G_6 + (2D-2)(3 D -4)(4 D -6)(5 D -8) (6 D -10)  \left(\Box+6 m^2\right) G_6 = 0\\
  \end{split}\end{equation*}

\item $n=7$:
  \begin{equation*}\begin{split} & x^6
   \left(\Box+m^2\right) \left(\Box+9 m^2\right) \left(\Box+25 m^2\right) \left(\Box+49
   m^2\right) G_7 +\\
   & 24 (D -1) x^4 \Lambda  \left(\Box^3+63 \Box^2 m^2+987 \Box m^4+3229 m^6\right) G_7
   +\\
   & 4 (D
   -1) x^4 \left(\Box^3 (61 D -72)+21 \Box^2 (133 D -156) m^2+3 \Box (9145 D
   -10656) m^4+7 (5257 D -6084) m^6\right) G_7 +\\
   & 12 (D -1) (3 D -4) x^2 \left(\Box^2 (D  (103 D -287)+200)+2 \Box (D (1021 D -2785)+1896) m^2+\right.\\
  & \qquad \left.7 (D  (721 D -1937)+1296) m^4\right) G_7 +\\
  & 24 (D -1) 4 (3 D -4) x^2
   \Lambda \left(\Box^2 (4 D -5)+7 \Box (17 D -21) m^2+(559 D -690) m^4\right) G_7 +\\
   & 24 (D -1) 2 (2 D -3) (3 D -5) (3 D -4) (5 D -8) \Lambda \left(\Box+9 m^2\right) G_7 +\\
   & 8 (D -1) (2 D -3) (3
   D -5) (3 D -4) (5 D -8) (7 D -12) \left(\Box+7 m^2\right) G_7 = 0\\
  \end{split}\end{equation*}

\item $n=8$:
   \begin{equation*}\begin{split} & x^6 \Lambda \left(\Box+4 m^2\right) \left(\Box+16 m^2\right) \left(\Box+36 m^2\right)
       \left(\Box+64 m^2\right) G_8+\\
       & 8 (D -1) x^6 \left(4
   \Box^4+375 \Box^3 m^2+9828 \Box^2 m^4+72160 \Box m^6+73728 m^8\right) G_8+\\
   & 4 (D -1) x^4 \Lambda \left(\Box^3 (103 D -120)+6 \Box^2 (1199 D -1392)
   m^2+12 \Box (10373 D -12009) m^4+64 (7009 D -8112) m^6\right) G_8 +\\
   & 8 (D -1) x^4 \left(\Box^3 (4 D -5) (97 D -120)+6 \Box^2 (D  (3301 D -8149)+5028) m^2+6 \Box (D  (36389 D
   -89133)+54558) m^4+\right.\\
   &\qquad \left. 256 (4 D -5) (338 D -399) m^6\right) G_8 +\\
   & 12 (D -1) (3 D -4) x^2 \Lambda \left(\Box^2 (D  (327 D -887)+600)+4 \Box
   (D  (2687 D -7223)+4836) m^2+\right.\\
   &\qquad \left. 4 (D  (13961 D -37509)+25098)
   m^4\right) G_8 +\\
   & 8 (D -1) 6 (2 D -3) (3 D -4) x^2 \left(2 \Box^2 (D  (59 D -168)+120)+\Box (D  (2623 D -7257)+5024) m^2+\right.\\
   &\qquad\left. 32
   (4 D  (59 D -160)+433) m^4\right) G_8 +\\
   &56 (D -1)
   (2 D -3) (3 D -5) (3 D -4) (5 D -8) (7 D -12) \Lambda \left(\Box+10
   m^2\right) G_8 + \\
   & 8 (D -1) 2 (2 D -3) (3 D
   -5) (3 D -4) (4 D -7) (5 D -8) (7 D -12) \left(\Box+8
   m^2\right)G_8 = 0\\
   \end{split}\end{equation*}
   
\item $n=9$:
  \begin{equation*}\begin{split} & x^8 \left(m^2+\Box\right) \left(9 m^2+\Box\right)
      \left(25 m^2+\Box\right) \left(49 m^2+\Box\right) \left(81 m^2+\Box\right) G_9 +\\
      & x^6 \Lambda \left(1057221 m^8+345620 m^6 \Box+26334 m^4
      \Box^2+660 m^2 \Box^3+5 \Box^4\right) G_9 +\\
      & 4 (D -1) x^6 \left(81 (62631 D -70424) m^8+44 (96599 D
      -109132) m^6 \Box+66 (7913 D -8976) m^4 \Box^2+\right.\\
      &\qquad \left. 132 (137 D -156) m^2 \Box^3+25 (7
   D -8) \Box^4\right) G_9 +\\
   & 8 (4 D -5) x^4 \Lambda
   \left(11 (12952 D -15291) m^6+33 (1097 D -1296) m^4 \Box+33 (58 D -69)
   m^2 \Box^2+5 (5 D -6) \Box^3\right) G_9 +\\
   & 12 (D -1) (3 D -4) x^4 \left(9 (3 D 
   (51453 D -126523)+233344) m^6+11 (D  (68903 D -171577)+106872) m^4
   \Box+\right.\\
      &\qquad \left. 11 (D  (5639 D -14225)+8976) m^2 \Box^2+(D  (1103 D -2817)+1800)
   \Box^3\right) G_9  +\\
   & 6 (2 D -3) (3
   D -4) x^2 \Lambda \left(11 (D  (8279 D -22453)+15192) m^4+\right.\\
      &\qquad \left. 22 (D  (727 D
   -1973)+1336) m^2 \Box+(5 D -7) (89 D -120) \Box^2\right) G_9 +\\
   & 8 (D -1) (2 D -3) (3 D -4)
   x^2 \left(27 (9 D  (D  (2369 D -10171)+14528)-62144) m^4+\right.\\
      &\qquad \left. 22 (D 
   (D  (7953 D -34681)+50410)-24432) m^2 \Box+(D  (23 D  (309 D
   -1379)+47260)-23520) \Box^2\right) G_9 +\\
   & 8 (D -1) \Lambda 16 (2 D -3) (3 D -5) (3 D
   -4) (4 D -7) (5 D -8) (7 D -12) \left(11 m^2+\Box\right) G_9 + \\
   & 16 (D -1) (2 D -3) (3 D -5) (3 D -4) (4 D -7) (5 D -8) (7 D
   -12) (9 D -16) \left(9 m^2+\Box\right) G_9 = 0\\
  \end{split}\end{equation*}

\end{itemize}
}
\subsection{Generic masses}
{\footnotesize
\begin{itemize}
\item $n=2$
  \begin{equation*}
  x^2\left(\Box + (m_1+m_2)^2\right)\left(\Box + (m_1-m_2)^2\right)G_2
  + 2(D-1)\Lambda\left(\Box + m_1^2+m_2^2\right)G_2 + 2(D-1)(D-2)\left(\Box + m_1^2+m_2^2\right) G_2 = 0
  \end{equation*}

\item $n=3$
    \begin{equation*}
    \begin{split}
      & 16 x^{10} \left(\left(m_1+m_2-m_3\right){}^2+\Box\right) \left(m_1-m_2-m_3\right)
   \left(m_1+m_2-m_3\right) \left(m_1-m_2+m_3\right) \left(m_1+m_2+m_3\right)\\
   &\qquad
   \left(\left(m_1-m_2+m_3\right){}^2+\Box\right)
   \left(\left(-m_1+m_2+m_3\right){}^2+\Box\right)
   \left(\left(m_1+m_2+m_3\right){}^2+\Box\right) G_3 +\\
   & x^8 \Lambda  \left(128 (D-1)
   \left(m_1-m_2-m_3\right) \left(m_1+m_2-m_3\right) \left(m_1-m_2+m_3\right)
   \left(m_1+m_2+m_3\right) \Box^3+\right. \\
   &\qquad \left. 384 (D-1) \left(m_1-m_2-m_3\right)
   \left(m_1+m_2-m_3\right) \left(m_1-m_2+m_3\right) \left(m_1+m_2+m_3\right)
   \left(m_1^2+m_2^2+m_3^2\right) \Box^2+\right. \\
   &\qquad \left. 128 (D-1) \left(m_1-m_2-m_3\right)
   \left(m_1+m_2-m_3\right) \left(m_1-m_2+m_3\right) \left(m_1+m_2+m_3\right)\right. \\
   &\qquad\qquad \left.
   \left(3 m_1^4+2 m_2^2 m_1^2+2 m_3^2 m_1^2+3 m_2^4+3 m_3^4+2 m_2^2 m_3^2\right)
   \Box+\right. \\
   &\qquad \left. 128 (D-1) \left(m_1-m_2-m_3\right) \left(m_1+m_2-m_3\right)
   \left(m_1-m_2+m_3\right) \left(m_1+m_2+m_3\right) \right. \\
   &\qquad\qquad \left. \left(m_1^6-m_2^2 m_1^4-m_3^2
   m_1^4-m_2^4 m_1^2-m_3^4 m_1^2+10 m_2^2 m_3^2 m_1^2+m_2^6+m_3^6-m_2^2
   m_3^4-m_2^4 m_3^2\right)\right) G_3 +\\
   & 8 (D-1)  x^8 \left(3 (7 D-13) m_1^{10}+\left(9
   (13-7 D) m_2^2+9 (13-7 D) m_3^2+8 \Box (13 D-24)\right) m_1^8+\right. \\
   &\qquad \left. 2 \left(3 (7 D-13)
   m_2^4+2 \left((51 D-81) m_3^2+4 \Box (17-9 D)\right) m_2^2+\right. \right. \\
   &\qquad\qquad \left.\left.3 (7 D-13) m_3^4+8 \Box
   (17-9 D) m_3^2+\Box^2 (73 D-135)\right) m_1^6+\right. \\
   &\qquad \left. 2 \left(3 (7 D-13) m_2^6+\left(3
   (69-47 D) m_3^2+40 \Box (D-2)\right) m_2^4+\right. \right. \\
   &\qquad\qquad \left.\left. \left(3 (69-47 D) m_3^4+24 \Box (11-7 D)
   m_3^2+\Box^2 (111-65 D)\right) m_2^2+3 (7 D-13) m_3^6+\right. \right. \\
   &\qquad\qquad \left.\left. 40 \Box (D-2) m_3^4+\Box^2 (111-65
   D) m_3^2+4 \Box^3 (8 D-15)\right) m_1^4+\right. \\
   &\qquad \left. \left(9 (13-7 D) m_2^8+16 \Box (17-9 D)
   m_2^6+2 \Box^2 (111-65 D) m_2^4+16 \Box^3 (12-7 D) m_2^2+9 (13-7 D) m_3^8+\right. \right. \\
   &\qquad\qquad \left.\left. 4 \left((51
   D-81) m_2^2+4 \Box (17-9 D)\right) m_3^6+2 \left(3 (69-47 D) m_2^4+24 \Box (11-7 D)
   m_2^2+\Box^2 (111-65 D)\right) m_3^4+\right. \right. \\
   &\qquad\qquad \left.\left. 4 \left((51 D-81) m_2^6+12 \Box (11-7 D) m_2^4+9
   \Box^2 (41-23 D) m_2^2+4 \Box^3 (12-7 D)\right) m_3^2+\Box^4 (D-3)\right) m_1^2+\right. \\
   &\qquad \left. 3 (7
   D-13) m_2^{10}+m_2^8 \left(9 (13-7 D) m_3^2+8 \Box (13 D-24)\right)+\right. \\
   &\qquad \left. 2 m_2^6
   \left(3 (7 D-13) m_3^4+8 \Box (17-9 D) m_3^2+\Box^2 (73 D-135)\right)+\right. \\
   &\qquad \left.m_3^2
   \left(m_3^2+\Box\right){}^2 \left(3 (7 D-13) m_3^4+2 \Box (31 D-57) m_3^2+\Box^2
   (D-3)\right)+\right. \\
   &\qquad \left. 2 m_2^4 \left(3 (7 D-13) m_3^6+40 \Box (D-2) m_3^4+\Box^2 (111-65 D)
   m_3^2+4 \Box^3 (8 D-15)\right)+\right. \\
   &\qquad \left. m_2^2 \left(9 (13-7 D) m_3^8+16 \Box (17-9 D) m_3^6+2
   \Box^2 (111-65 D) m_3^4+16 \Box^3 (12-7 D) m_3^2+\Box^4 (D-3)\right)\right) G_3+\\
   & x^6 \Lambda 
   \left(16 (D-3) (D-1) (4 D-3) \left(m_1^2+m_2^2+m_3^2\right) \Box^3+\right. \\
   &\qquad \left. 16 (D-3) (D-1)
   \left(56 D m_1^4-87 m_1^4-64 D m_2^2 m_1^2+142 m_2^2 m_1^2-64 D m_3^2 m_1^2+142
   m_3^2 m_1^2+56 D m_2^4-87 m_2^4+\right. \right. \\
   &\qquad\qquad \left.\left. 56 D m_3^4-87 m_3^4-64 D m_2^2 m_3^2+142 m_2^2
   m_3^2\right) \Box^2+\right. \\
   &\qquad \left. 16 (D-3) (D-1) \left(m_1^2+m_2^2+m_3^2\right) \left(84 D
   m_1^4-141 m_1^4-136 D m_2^2 m_1^2+266 m_2^2 m_1^2-136 D m_3^2 m_1^2+266 m_3^2
   m_1^2+\right.\right. \\
   &\qquad \qquad\left.\left. 84 D m_2^4-141 m_2^4+84 D m_3^4-141 m_3^4-136 D m_2^2 m_3^2+266 m_2^2
   m_3^2\right) \Box+\right. \\
   &\qquad \left.16 (D-3) (D-1) \left(32 D m_1^8-57 m_1^8-48 D m_2^2 m_1^6+108
   m_2^2 m_1^6-48 D m_3^2 m_1^6+108 m_3^2 m_1^6+32 D m_2^4 m_1^4-102 m_2^4
   m_1^4+\right.\right. \\
   &\qquad \qquad\left.\left. 32 D m_3^4 m_1^4-102 m_3^4 m_1^4-48 D m_2^2 m_3^2 m_1^4+132 m_2^2 m_3^2
   m_1^4-48 D m_2^6 m_1^2+108 m_2^6 m_1^2-48 D m_3^6 m_1^2+108 m_3^6 m_1^2-\right.\right. \\
   &\qquad \qquad\left.\left. 48 D
   m_2^2 m_3^4 m_1^2+132 m_2^2 m_3^4 m_1^2-48 D m_2^4 m_3^2 m_1^2+132 m_2^4 m_3^2
   m_1^2+32 D m_2^8-57 m_2^8+32 D m_3^8-57 m_3^8-\right.\right. \\
   &\qquad \qquad\left.\left. 48 D m_2^2 m_3^6+108 m_2^2
   m_3^6+32 D m_2^4 m_3^4-102 m_2^4 m_3^4-48 D m_2^6 m_3^2+108 m_2^6
   m_3^2\right)\right) G_3 -\\
   & (D-3) (D-1) x^6 \left(-3 (7 D (25 D-92)+685) m_1^8-\right. \\
   &\qquad \left. 4
   \left(-3 (3 D (21 D-92)+365) m_2^2-3 (3 D (21 D-92)+365) m_3^2+\Box (D (481
   D-1882)+1953)\right) m_1^6+\right. \\
   &\qquad \left. \left(-6 (D (77 D-460)+775) m_2^4+4 \left(3 (D (65
   D-332)+243) m_3^2+\Box (D (225 D-1402)+1729)\right) m_2^2-\right.\right. \\
   &\qquad \qquad\left.\left. 6 (D (77 D-460)+775)
   m_3^4+4 \Box (D (225 D-1402)+1729) m_3^2-2 \Box^2 (D (815 D-3056)+3129)\right)
   m_1^4+\right. \\
   &\qquad \left. 4 \left(3 (3 D (21 D-92)+365) m_2^6+\left(3 (D (65 D-332)+243) m_3^2+\Box (D
   (225 D-1402)+1729)\right) m_2^4+\right.\right. \\
   &\qquad \qquad\left.\left. \left(3 (D (65 D-332)+243) m_3^4+6 \Box (D (361
   D-1674)+1673) m_3^2+\Box^2 (D (267 D-1888)+2221)\right) m_2^2+\right.\right. \\
   &\qquad \qquad\left.\left. 3 (3 D (21
   D-92)+365) m_3^6+\Box (D (225 D-1402)+1729) m_3^4+\right.\right. \\
   &\qquad \qquad\left.\left. \Box^2 (D (267 D-1888)+2221)
   m_3^2-3 \Box^3 (D (19 D-42)+47)\right) m_1^2-\right. \\
   &\qquad \left. 3 (7 D (25 D-92)+685) m_2^8-4 m_2^6
   \left(\Box (D (481 D-1882)+1953)-3 (3 D (21 D-92)+365) m_3^2\right)-\right. \\
   &\qquad \left. 2 m_2^4
   \left(3 (D (77 D-460)+775) m_3^4-2 \Box (D (225 D-1402)+1729) m_3^2+\Box^2 (D (815
   D-3056)+3129)\right)+\right. \\
   &\qquad \left. 4 m_2^2 \left(3 (3 D (21 D-92)+365) m_3^6+\Box (D (225
   D-1402)+1729) m_3^4+\right.\right. \\
   &\qquad \qquad\left.\left. \Box^2 (D (267 D-1888)+2221) m_3^2-3 \Box^3 (D (19
   D-42)+47)\right)+\right. \\
   &\qquad \left. \left(m_3^2+\Box\right) \left(-3 (7 D (25 D-92)+685) m_3^6+\Box
   (-1399 (D-4) D-5757) m_3^4-\right.\right. \\
   &\qquad \qquad\left.\left. 3 \Box^2 (D (77 D-172)+167) m_3^2+3 \Box^3 (D-7)
   (D+3)\right)\right) G_3 +\\
   \end{split}\end{equation*}
  \begin{equation*}\begin{split}
   & x^4 \Lambda  \left(-12 (D-7) (D-3) (D-1) (D+3) (2 D-1)
   \Box^3+\right. \\
   &\qquad \left. 4 (D-3) (D-1) \left(70 D^3-357 D^2+1034 D-459\right)
   \left(m_1^2+m_2^2+m_3^2\right) \Box^2+\right. \\
   &\qquad \left. 4 (D-3) (D-1) \left(222 D^3 m_1^4-1587 D^2
   m_1^4+4446 D m_1^4-3201 m_1^4+84 D^3 m_2^2 m_1^2+574 D^2 m_2^2 m_1^2-3564 D
   m_2^2 m_1^2+\right.\right. \\
   &\qquad \qquad\left.\left. 4298 m_2^2 m_1^2+84 D^3 m_3^2 m_1^2+574 D^2 m_3^2 m_1^2-3564 D
   m_3^2 m_1^2+4298 m_3^2 m_1^2+222 D^3 m_2^4-1587 D^2 m_2^4+\right.\right. \\
   &\qquad \qquad\left.\left. 4446 D m_2^4-3201
   m_2^4+222 D^3 m_3^4-1587 D^2 m_3^4+4446 D m_3^4-3201 m_3^4+84 D^3 m_2^2
   m_3^2+574 D^2 m_2^2 m_3^2-\right.\right. \\
   &\qquad \qquad\left.\left. 3564 D m_2^2 m_3^2+4298 m_2^2 m_3^2\right) \Box+\right. \\
   &\qquad \left. 4 (D-3)
   (D-1) \left(146 D^3 m_1^6-1203 D^2 m_1^6+3526 D m_1^6-2805 m_1^6-18 D^3 m_2^2
   m_1^4+627 D^2 m_2^2 m_1^4-2694 D m_2^2 m_1^4+\right.\right. \\
   &\qquad \qquad\left.\left. 2421 m_2^2 m_1^4-18 D^3 m_3^2
   m_1^4+627 D^2 m_3^2 m_1^4-2694 D m_3^2 m_1^4+2421 m_3^2 m_1^4-18 D^3 m_2^4
   m_1^2+627 D^2 m_2^4 m_1^2-\right.\right. \\
   &\qquad \qquad\left.\left. 2694 D m_2^4 m_1^2+2421 m_2^4 m_1^2-18 D^3 m_3^4
   m_1^2+627 D^2 m_3^4 m_1^2-2694 D m_3^4 m_1^2+2421 m_3^4 m_1^2-588 D^3 m_2^2
   m_3^2 m_1^2+\right.\right. \\
   &\qquad \qquad\left.\left. 5778 D^2 m_2^2 m_3^2 m_1^2-16260 D m_2^2 m_3^2 m_1^2+17694 m_2^2
   m_3^2 m_1^2+146 D^3 m_2^6-1203 D^2 m_2^6+3526 D m_2^6-2805 m_2^6+\right.\right. \\
   &\qquad \qquad\left.\left. 146 D^3
   m_3^6-1203 D^2 m_3^6+3526 D m_3^6-2805 m_3^6-18 D^3 m_2^2 m_3^4+627 D^2 m_2^2
   m_3^4-2694 D m_2^2 m_3^4+2421 m_2^2 m_3^4-\right.\right. \\
   &\qquad \qquad\left.\left. 18 D^3 m_2^4 m_3^2+627 D^2 m_2^4
   m_3^2-2694 D m_2^4 m_3^2+2421 m_2^4 m_3^2\right)\right) G_3 -\\
   & 2 (D-3) (D-1)
   x^4 \left((D (D ((2993-289 D) D-12011)+19567)-11220) m_1^6+\right. \\
   &\qquad \left. 3 \left((D (D (D (11
   D-443)+2681)-5157)+3228) m_2^2+(D (D (D (11 D-443)+2681)-5157)+3228) m_3^2+\right.\right. \\
   &\qquad \qquad\left.\left. \Box (D
   (D ((1575-167 D) D-6189)+9257)-4796)\right) m_1^4+\right. \\
   &\qquad \left. \left(3 (D (D (D (11
   D-443)+2681)-5157)+3228) m_2^4+\right.\right. \\
   &\qquad \qquad\left.\left. 6 (D (D (D (197
   D-2305)+9235)-17375)+11208) m_3^2 m_2^2+\right.\right.\\
   &\qquad \qquad \left.\left.2 \Box (D (D ((407-171 D)
   D+1883)-8423)+7264)m_2^2 +\right.\right. \\
   &\qquad \qquad\left.\left. 3 (D (D (D (11 D-443)+2681)-5157)+3228)
   m_3^4+2 \Box (D (D ((407-171 D) D+1883)-8423)+7264) m_3^2-\right.\right. \\
   &\qquad \qquad \left. \left.\Box^2 (D-1) (D (D (167
   D-1304)+5837)-3924)\right) m_1^2+\right. \\
   &\qquad \left. (D (D ((2993-289 D) D-12011)+19567)-11220)
   m_2^6+\right. \\
   &\qquad\left.(D (D ((2993-289 D) D-12011)+19567)-11220) m_3^6-3 \Box (D (D (D (167
   D-1575)+6189)-9257)+4796) m_3^4-\right. \\
   &\qquad \left. \Box^2 (D-1) (D (D (167 D-1304)+5837)-3924)
   m_3^2+9 \Box^3 (D-7) (D-1) (D+3) (5 D-4)+\right. \\
   &\qquad \left. 3 m_2^4 \left((D (D (D (11
   D-443)+2681)-5157)+3228) m_3^2+\Box (D (D ((1575-167 D)
   D-6189)+9257)-4796)\right)+\right. \\
   &\qquad \left. m_2^2 \left(3 (D (D (D (11 D-443)+2681)-5157)+3228)
   m_3^4+2 \Box (D (D ((407-171 D) D+1883)-8423)+7264) m_3^2-\right.\right. \\
   &\qquad \qquad\left.\left. \Box^2 (D-1) (D (D (167
   D-1304)+5837)-3924)\right)\right) G_3 -\\
  \end{split}\end{equation*}
  \begin{equation*}\begin{split}
   & 12 (D-3) (D-2) (D-1)^2 x^2 \left(-(D-2) (D (D
   (9 D-124)+619)-528) m_1^4-\right. \\
   &\qquad \left. 2 (D-2) (3 D-4) (D (5 D-44)+111) m_2^2 m_1^2-2 (D-2)
   (3 D-4) (D (5 D-44)+111) m_3^2 m_1^2 +\right. \\
   &\qquad \left. 2 \Box (D (D (D (4 D+7)-468)+1417)-1032)m_1^2-(D-2) (D (D (9 D-124)+619)-528) m_2^4+\right. \\
   &\qquad \left. 2 m_2^2 \left(\Box (D (D (D (4
   D+7)-468)+1417)-1032)-(D-2) (3 D-4) (D (5 D-44)+111)
   m_3^2\right)+\right. \\
   &\qquad \left. \left(m_3^2+\Box\right) \left(\Box (D-7) (D+3) (D (17 D-60)+48)-(D-2) (D
   (D (9 D-124)+619)-528) m_3^2\right)\right) G_3 +\\
   & x^2 \Lambda  \left(-12 \Box^2 (D-7)
   (D-3) (D-2) (D+3) (11 D-12) (D-1)^2-\right. \\
   &\qquad \left. 24 \Box (D-3) (D-2) \left(D^3+34 D^2-401
   D+390\right) \left(m_1^2+m_2^2+m_3^2\right) (D-1)^2+\right. \\
   &\qquad \left. 12 (D-3) (D-2) \left(9 D^3
   m_1^4-124 D^2 m_1^4+619 D m_1^4-528 m_1^4+30 D^3 m_2^2 m_1^2-304 D^2 m_2^2
   m_1^2+1018 D m_2^2 m_1^2-\right. \right. \\
   &\qquad \left.\left. 888 m_2^2 m_1^2+30 D^3 m_3^2 m_1^2-304 D^2 m_3^2
   m_1^2+1018 D m_3^2 m_1^2-888 m_3^2 m_1^2+9 D^3 m_2^4-124 D^2 m_2^4+619 D
   m_2^4-528 m_2^4+\right. \right. \\
   &\qquad\qquad \left. \left.9 D^3 m_3^4-124 D^2 m_3^4+619 D m_3^4-528 m_3^4+30 D^3 m_2^2
   m_3^2-304 D^2 m_2^2 m_3^2+1018 D m_2^2 m_3^2-888 m_2^2 m_3^2\right)
   (D-1)^2\right) G_3 -\\
   &24 (D-7) (D-4) (D-3)^2 (D-2)^2 (D-1)^2 (D+3) (3 D-4)
   \left(m_1^2+m_2^2+m_3^2+\Box\right) G_3-\\
   & 24 (D-7) (D-4) (D-3)^2 (D-2) (D-1)^2 (D+3) (3
   D-4) \Lambda  \left(m_1^2+m_2^2+m_3^2+\Box\right) G_3 = 0\\
    \end{split}
  \end{equation*}

\end{itemize}
}

\end{document}